\documentclass[a4paper,11pt]{article}

\usepackage{jheppub}  
\usepackage[font=footnotesize,labelfont=bf]{caption}
\usepackage{subcaption}
\usepackage{graphicx}
\usepackage{bm}
\usepackage{xcolor}
\usepackage{multirow}
\usepackage{enumerate}
\usepackage[normalem]{ulem}

\newcommand{\dd}{\mathrm{d}}
\newcommand{\Mpl}{M_{\text{pl}}}

\DeclareMathOperator\erf{erf}

\title{Gravitational waves from primordial black hole dominance: The effect of inflaton decay rate}

\author[a,b]{Daniel del-Corral}
\author[c]{K. Sravan Kumar}
\author[a,b]{Jo\~ao Marto}

\affiliation[a]{Departamento de F\'{\i}sica, Universidade da Beira Interior, Rua Marqu\^{e}s D'\'Avila e Bolama 6200-001 Covilh\~a, Portugal}

\affiliation[b]{Centro de Matem\'atica e Aplica\c{c}\~oes da Universidade da Beira Interior, Rua Marqu\^{e}s D'\'Avila e Bolama 6200-001 Covilh\~a, Portugal}

\affiliation[c]{Institute of Cosmology and Gravitation, University of Portsmouth,
Dennis Sciama Building, Burnaby Road,
Portsmouth, PO1 3FX, United Kingdom}

\emailAdd{corral.martinez@ubi.pt}
\emailAdd{sravan.kumar@port.ac.uk}
\emailAdd{jmarto@ubi.pt}

\abstract{
In this work, we explore primordial black holes (PBH) formation scenario during the post-inflationary preheating stage dominated by the inflaton field. We consider, in particular, a model-independent parametrization of the Gaussian peak inflationary power spectrum that leads to amplified inflationary density fluctuations before the end of inflation. These modes can reenter the horizon during preheating and could experience instabilities that trigger the production of PBH. This is estimated with the Khlopov-Polnarev (KP) formalism that takes into account non-spherical effects. We derive an accurate analytical expression for the mass fraction under the KP formalism that fits well with the numerical evaluation.  Particularly, we focus on ultra-light PBH of masses $M_{\text{PBH}}<10^9g$ and study their evolution and (possible) dominance after the decay of the inflation field into radiation and before the PBH evaporation via Hawking radiation. These considerations alter the previous estimates of induced gravitational waves (GWs) from PBH dominance and set new targets for detecting stochastic GW backgrounds with future detectors, provided that these achieve significantly enhanced experimental sensitivity, as current planned instruments do not yet possess sufficient sensitivity for detection.}

\begin{document} 

\maketitle

\flushbottom

%%%%%%%%%%%%%%%%%%%%%
%% INTRODUCTION
%%%%%%%%%%%%%%%%%%%%%

\section{Introduction}

The inflationary phase constitutes the most important phase of the primordial Universe. It solves the conceptual challenges of the classical Big Bang Theory and provides a mechanism for generating the primordial density fluctuations observed in the cosmic microwave background (CMB). These fluctuations act as seeds for many phenomena, including primordial black holes (PBH) \cite{Zeldovich:1967lct,Hawking:1971ei,Carr:1974nx,Carr:1975qj} and gravitational waves (GWs) \cite{Guzzetti:2016mkm,Baumann:2009ds,Maggiore:2007,Dodelson:2003,Gravitation:1973,Franciolini:2021,Domenech:2021ztg,Baumann:2007,Ananda:2006,Assadullahi:2009,Kohri:2018awv}, in posterior eras. The specific nature of each era (matter or radiation-dominated) influences the dynamics of these phenomena, providing insights into early universe physics.

Almost every inflationary model, especially single-field models, exhibits a phase known as preheating \cite{Kofman:1994rk,Kofman:1997yn}, just after the end of inflation. During this stage, the universe undergoes a transient matter-dominated period driven by the oscillations of the inflaton field at the bottom of the inflationary potential. PBH are usually considered to form from the collapse of amplified density perturbations generated during inflation as they enter the horizon. Conventionally, PBH formation has been studied in radiation-dominated eras, where the radiation pressure allows the definition of a collapse threshold for the perturbations \cite{Carr:2020gox,Harada:2013epa,Musco:2012au}, and the mass fraction of PBH is often estimated using the Press-Schechter formalism \cite{Press:1973iz,Harada:2013epa}. In contrast, during this preheating stage, characterized by negligible pressure, the collapse dynamics differ significantly. Here, non-spherical effects become the dominant factor in stopping the collapse, a scenario well described by the Khlopov-Polnarev (KP) formalism \cite{Khlopov:1980mg,Polnarev:1981,Khlopov:1982ef,Polnarev:1985btg,Khlopov:2008qy,Harada:2016mhb}. This last, typically ignored in the literature, was originally thought for matter-dominated scenarios. We choose to work under this formalism as we understand that it better describes the collapse dynamics of perturbations in matter-dominated scenarios. We have already explored this in previous works \cite{del-Corral:2023apl,del-Corral:2025lcp}, but here, we consider two new elements: the decay of the inflation field into radiation and the posterior evaporation of the PBH via Hawking radiation, which we describe in what follows. Further, we will apply these considerations to an extended distribution of perturbations, contrary to the standard monochromatic assumption.

Since PBH form from the collapse of amplified perturbations, a central element in its formation is the appearance of a pronounced peak in the inflationary power spectrum. Various inflationary scenarios can lead to such an amplification of curvature perturbations. For instance, in single-field models, mechanisms such as an inflection point \cite{Iacconi:2021ltm,Dalianis:2018frf,Garcia-Bellido:2017mdw,Hertzberg:2017dkh} or a step-like feature in the potential \cite{Inomata:2021tpx,Dalianis:2023pur,Dalianis:2021iig,Cai:2021zsp,Papanikolaou:2022did} can generate the necessary peak. For multi-field, this peak can be produced through non-canonical kinetic terms \cite{Aldabergenov:2022rfc,Braglia:2020eai,Pi:2021dft,Wang:2024vfv}, the interplay of multiple axion fields \cite{Zhou:2020kkf,Mavromatos:2022yql}, waterfall trajectories in hybrid models \cite{Afzal:2024xci,Spanos:2021hpk,Braglia:2022phb,Clesse:2015wea,Tada:2023fvd} {or rapid turns in the inflationary trajectory \cite{Fumagalli:2020adf,Palma:2020ejf,Anguelova:2020nzl}}. For a comprehensive review of these mechanisms, see \cite{Stamou:2024lqf}. Although the parametrization of the power spectrum via a Gaussian peak is introduced in an ad hoc manner in our study, adjusting its height and position effectively encapsulates the essential physics of almost every inflationary model mentioned earlier. This is the approach we will follow, remaining agnostic about the precise physical mechanism of the amplification of perturbations during inflation.

Interest in PBH has grown recently due to their potential to answer several questions in cosmology \cite{Stojkovic:2004hz,Stojkovic:2005zh}. In particular, they have been proposed as candidates for dark matter, as generators of structure in the
universe, or even as seeds for the formation of supermassive black holes in the center of
galactic nuclei. See \cite{Villanueva-Domingo:2021spv,Carr:2020gox} for a review. However, although observational data tightly constrain their abundance \cite{Carr:2020gox,Cai:2021zsp,Markov:1984xd,Coleman:1991ku,Carr:1994ar,Zeldovich,Acharya:2020jbv,Carr:2009jm,Josan:2009qn,Carr:2020xqk}, for small masses ($\lesssim10^{10}$g), these constraints rely on the nature of Dark Matter, which is currently unknown \cite{Allahverdi:2020bys}. For this reason, the constraints can be relaxed to the point that PBH come to dominate the energy density of the universe, a period called PBH-dominated, which behaves as pressureless matter. If this occurs, their inherent Poissonian fluctuations in density can source a stochastic background of gravitational waves, providing a unique observational window into the early universe. {The frequencies associated with this background of GWs fall into the very-high frequency range, as shown in previous works in the context of PBH in grand unified theories and supergravity \cite{Anantua:2008am,Zagorac:2019ekv,del-Corral:2025fzz}. In this range, the current and planned detectors still lack a sufficiently high sensitivity, further motivating their refinement.}

To eventually recover the standard radiation-dominated era necessary for successful Big Bang Nucleosynthesis (BBN), we consider the inflaton decay into Standard Model particles, thereby reheating the universe. This is achieved by considering the scalar field's decay into radiation, with decay rate $\Gamma_\phi$. However, if PBH dominates before that, then the universe is reheated through Hawking evaporation \cite{Hawking:1974rv,Martin:2019nuw,Gondolo:2020uqv,Dong:2015yjs,Holst:2024ubt}, and the reheating temperature corresponds to the evaporation temperature of the PBH. Refs.~\cite{Papanikolaou:2020qtd,Domenech:2023jve,Domenech:2024wao,Domenech:2020ssp,Papanikolaou:2022chm,Papanikolaou:2024kjb,He:2024luf,Inomata:2019ivs} considered that the PBH-dominated phase occurs due to the interplay between the different energy densities of PBH (matter), and a background fluid with an arbitrary equation of state $w>0$, ignoring the specific formation mechanism of the PBH. We, however, consider that PBH are produced during 
an early matter-dominated phase, under the KP formalism, and that the PBH-dominated phase occurs due to the decay of the scalar field into radiation, which modifies the actual constraints on the abundance of PBH. Then, we analyze the production of induced GWs by the Poissonian fluctuations of the PBH fluid. To our knowledge, this is the first time it has been studied in the literature.

This work is organized as follows: We begin in Sec.~\ref{sec:inflation-and-preheating} by defining an analytical expression for our power spectrum and giving some details on the preheating period and the parameters of the model. Then, in Sec.~\ref{sec:PBH-dominated}, we describe the dynamics of a PBH-dominated phase, from the collapse of perturbations to the power spectrum of the PBH density fluctuations. The results for the induced GWs are obtained in Sec.~\ref{sec:IGWs}, as well as a comparison with previous studies in the literature. Conclusions are given in Sec.~\ref{sec:conclusions}, and appendices \ref{sec:appendix}, \ref{sec:appendix2}, and \ref{sec:appendix3} show some analytical estimations of the mass fraction of PBH and the fractional energy densities of PBH and GWs, respectively. Throughout the paper, we consider mostly positive metric signature $\left( -+++ \right)$, set $\hbar=c=1$ and consider the units of reduced Planck mass $M_p^2 = \frac{1}{8\pi G} = 2.4\times 10^{18}\rm GeV$.

%%%%%%%%%%%%%%%%%%%%%%%%%%%%
%% INFLATIONARY PARAMETERS
%%%%%%%%%%%%%%%%%%%%%%%%%%%%

\section{Inflation and preheating}\label{sec:inflation-and-preheating}

\begin{figure}
    \centering
    \includegraphics[width=0.75\linewidth]{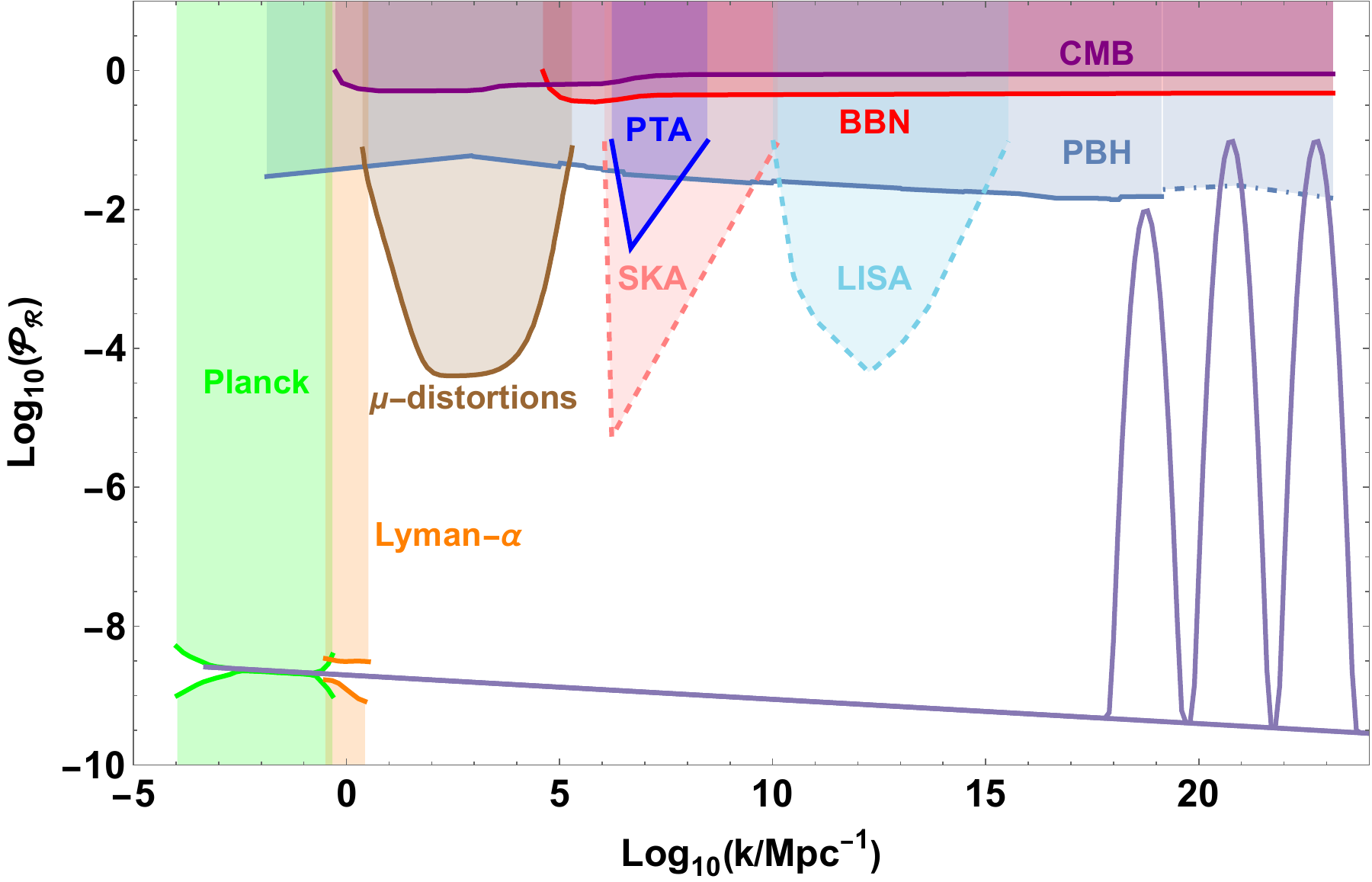}
    \caption{Three examples of inflationary power spectra + gaussian peak (corresponding to $A_{\rm peak} = 0.01,\,0.1,\,0.1$ and $k_{\rm peak} = 6\times 10^{18}, 6\times 10^{20}, 6\times 10^{22} {\rm Mpc}^{-1}$ respectively), that we compare against the constraints on the power spectrum of curvature perturbations from PBH \cite{Josan:2009qn,Kalaja:2019uju,Carr:2020gox}, Planck observations on the CMB \cite{Planck:2018vyg}, Lyman-$\alpha$ forest \cite{Murgia:2019duy}, $\mu$-distortions \cite{Chluba:2012we} and GWs \cite{Byrnes:2018txb,Mahbub:2021qeo,Allahverdi:2020bys} (PTA, SKA and LISA). The continuous (dashed) lines represent current (forecasted) constraints. The dotted-dashed lines for the PBH constraint represent the constraints associated with the non-standard scenarios like Hawking evaporation considerations and the possibility of PBH remnants. The above Figure is produced from the data available in \cite{Josan:2009qn,Mahbub:2021qeo}.}
    \label{fig:P-constraints}
\end{figure}

As stated in the introduction, we consider an inflationary power spectrum with a Gaussian peak. According to Planck's 2018 results \cite{Planck:2018vyg,Planck:2018jri}, the inflationary power spectrum of curvature perturbations $\mathcal R_{\bm k}$ can be parametrized as follows
\begin{equation}\label{eq:power-spectrum-inf}
    \mathcal{P}_{\mathcal{R}}^{\text{inf}}(k)=\mathcal{A}_s\left(\frac{k}{k_0}\right)^{n_s-1},
\end{equation}
where $\mathcal{A}_s=2.1\times10^{-9}$ is the amplitude of the power spectrum at the pivot scale $k_0=0.05\text{Mpc}^{-1}$, and $n_s\simeq0.965$ is the spectral index. Motivated by other works \cite{Iacconi:2021ltm,Dalianis:2018frf,Stamou:2024lqf,Garcia-Bellido:2017mdw,Hertzberg:2017dkh,Inomata:2021tpx,Dalianis:2023pur,Dalianis:2021iig,Cai:2021zsp,Aldabergenov:2022rfc,Braglia:2020eai,Pi:2021dft,Wang:2024vfv,Zhou:2020kkf,Mavromatos:2022yql,Afzal:2024xci,Spanos:2021hpk,Braglia:2022phb,Clesse:2015wea,Tada:2023fvd,Papanikolaou:2022did,Fumagalli:2020adf,Palma:2020ejf,Anguelova:2020nzl} (see the introduction), we introduce a Gaussian peak in the power spectrum, and choose to parametrize it as a log-normal-like distribution, given by
\begin{equation}\label{eq:power-spectrum-peak}
    \mathcal{P}_{\mathcal{R}}^{\text{peak}}(k)=\mathcal{A}_{\text{peak}}\exp\left[-\frac{\left(\log_{10}(k/k_{\text{peak}})\right)^2}{\sigma^2}\right],
\end{equation}
where $\mathcal{A}_{\text{peak}}$ is the amplitude at the peak scale $k_{\text{peak}}$ and the standard deviation of the Gaussian is chosen to be $\sigma=0.2$ to reduce the number of free parameters. Eqn.~\eqref{eq:power-spectrum-peak} is useful for this work as it avoids the tails of the Gaussian peak (due to the $\log_{10}$ term), and since it constitutes a two-parameter ($\mathcal{A}_\text{peak}$ and $k_\text{peak}$) model-independent parametrization that can reflect the physics of the models described in \cite{Iacconi:2021ltm,Dalianis:2018frf,Garcia-Bellido:2017mdw,Hertzberg:2017dkh,Inomata:2021tpx,Dalianis:2023pur,Dalianis:2021iig,Cai:2021zsp,Aldabergenov:2022rfc,Braglia:2020eai,Pi:2021dft,Wang:2024vfv,Zhou:2020kkf,Mavromatos:2022yql,Afzal:2024xci,Spanos:2021hpk,Braglia:2022phb,Clesse:2015wea,Tada:2023fvd,Stamou:2024lqf,Papanikolaou:2022did,Fumagalli:2020adf,Palma:2020ejf,Anguelova:2020nzl}. The total power spectrum is the sum of the contributions from \eqref{eq:power-spectrum-inf} and \eqref{eq:power-spectrum-peak}, that is
\begin{equation} \label{eq:power-spectrum-total}
    \mathcal{P}_{\mathcal{R}}(k)=\mathcal{P}_{\mathcal{R}}^{\text{inf}}(k)+\mathcal{P}_{\mathcal{R}}^{\text{peak}}(k).
\end{equation}
In Fig.~\ref{fig:P-constraints} we show some examples of the power spectrum \eqref{eq:power-spectrum-total} along with several current (continuous) and forecasted (dashed) constraints on the power spectrum. The dotted-dashed constraints from PBH represent non-standard scenarios that depend on the detailed nature of Dark Matter, which is currently unknown, as well as on the details of the Hawking radiation and Planck-size remnants after PBH evaporation \cite{Allahverdi:2020bys}. We will remain agnostic about these last constraints and consider values of $
\mathcal{P}_{\mathcal{R}}(k)$ that are both above and below them, as illustrated in Fig.~\ref{fig:P-constraints}. Furthermore, the positions of the Gaussian peaks of the power spectra shown in Fig.~\ref{fig:P-constraints} correspond to scales that exit the horizon at a time close to the end of inflation, when the scalar field is of the order of the Planck mass.

According to Planck's results \cite{Planck:2018vyg,Planck:2018jri}, the upper limit on the energy density at the pivot scale is given by $\rho_{\text{inf}}\lesssim 10^{16}\text{GeV}$, which translates into the following upper limit on the Hubble rate at the pivot scale $H_{\text{inf}}\lesssim2.5\times10^{-5}\Mpl$. We further assume that $H$ is constant during inflation and extrapolate that value till the end of inflation, where the preheating period begins. This phase is characterized by an oscillating scalar field at the bottom of the inflationary potential, which makes the universe (effectively) behave as being matter-dominated and implies that the scale factor can be parametrized as
\begin{equation}\label{eq:scale-factor}
    a(t)\simeq a_{\text{end}}\left(\frac{t}{t_{\text{end}}}\right)^{2/3},
\end{equation}
where the suffix ``$_{\text{end}}$'' means the quantity at the end of inflation and $t$ is cosmic time. Also, the Hubble rate, defined as $H(t)=\dot a/a$, with an overdot denoting a derivative with respect to cosmic time, is given by
\begin{equation}\label{eq:hubble-rate}
    H(t)\simeq\frac2{3t}.
\end{equation}
During this phase, the perturbations that enter the horizon collapse into PBH, as shown in the next section. However, the quantity we use to compute the abundance of PBH, \textit{i.e.}, the mass fraction $\beta(k)$, is not the power spectrum of curvature perturbations, but the variance of the density perturbations, $\sigma_k$. For modes that enter the horizon after inflation, the curvature perturbations are related to the density perturbations $\delta_{\phi,\bm k}$ as follows \footnote{{In this work, since our focus is only on the scales that are amplified during inflation, we do not consider the subhorizon modes that never exit the horizon during inflation. Although they could contribute to PBH formation, as shown in \cite{del-Corral:2023apl,del-Corral:2025lcp}, their contribution is only toward very small scales in contrast to what we explore in this work.}}\cite{Martin:2019nuw,Jedamzik:2010dq}
\begin{equation}\label{eq:curvature-density-relation}
    \delta_{\phi,\bm k}\simeq\frac{8}{5}\mathcal R_{\bm k},
\end{equation}
where a suffix ``$_{\bm k}$'' means Fourier component and $k$ is the modulus of the wavenumber vector $\bm{k}$. Also, a suffix ``$_\phi$'' is introduced as a notation so that these fluctuations are not confused with the energy density fluctuation of the PBH, which is defined below. Eqn.~\eqref{eq:curvature-density-relation} implies that the power spectrum of the density perturbations should be written as
\begin{equation}\label{eq:power-spectrum-density}
    \mathcal P_{\delta_\phi}(k)\simeq\frac{64}{25}\mathcal P_\mathcal{R}(k),
\end{equation}
and we can approximate the variance as follows
\begin{equation}\label{eq:variance}
    \sigma_k\simeq\sqrt{\mathcal P_\mathcal{\delta_\phi}(k)}.
\end{equation}

As explained in the introduction, to terminate the preheating stage and recover the usual reheating scenario, the inflation field must decay into radiation via a decay rate $\Gamma_\phi$, which occurs at a time $t_\text{r}=\Gamma_\phi^{-1}$. At this point, the time evolution of the temperature of the radiation fluid is given by \cite{SravanKumar:2018tgk,Kofman:1994rk,Kofman:1997yn}
\begin{equation}\label{eq:T-decay}
    T_\text{decay}(t)=\left(\frac{90}{\pi^2g_*(T)}\right)^{1/4}\sqrt{\frac{\Gamma_\phi}{\Mpl}}\left(\frac{a_\text{r}}{a}\right)\Mpl,
\end{equation}
in units of Planck mass ($1\Mpl=1.22\times10^{19}\text{GeV}$). In this equation, $a_\text{r}=a(t_\text{r})$ and $g_*(T)$ is the number of relativistic degrees of freedom, which takes the value $g_*=106.75$ for $T\gtrsim100\text{GeV}$ (assuming the Standard Model is valid at those energies, see Refs.~\cite{Husdal:2016haj,Saikawa:2018rcs} for a review of the dependence of the effective degrees of freedom with temperature). However, if after the decay the abundance of PBH is high enough, they will dominate due to the different redshifts of the energy densities of radiation and PBH, $a^{-4}$ and $a^{-3}$, respectively. In this case, since radiation becomes a subdominant component, the reheating occurs at a later stage and through the evaporation of PBH due to Hawking radiation \cite{Hawking:1974rv}, with an associated temperature given by (see eqn.~(2.24) of \cite{Domenech:2020ssp})
\begin{equation}\label{eq:T-evap}
    T_\text{eva}\simeq2.76\times10^4\text{GeV}\left(\frac{g_*(T_\text{eva})}{106.75}\right)^{-1/4}\left(\frac{g_H}{108}\right)^{1/2}\left(\frac{M_{\text{PBH},0}}{10^4\text{g}}\right)^{-3/2},
\end{equation}
with $g_H$ being the spin-weighted degrees of freedom and $M_{\text{PBH},0}$ the mean mass of the PBH distribution, considered as the mass of the PBH associated with the Gaussian peak (see eqns.~\eqref{eq:PBH-mass} and \eqref{eq:mean-PBH-mass}).
Taking into account that for a successful BBN the temperature of the plasma should be $T_\text{BBN}>4\text{MeV}$ \cite{Hannestad:2004px,Kawasaki:2000en,Hasegawa:2019jsa} and the upper limit on the energy scale of inflation, $H_\text{inf}/\Mpl\lesssim2.5\times10^{-5}$ \cite{Planck:2018vyg,Planck:2018jri}, one obtains the following allowed range for the PBH masses, given by \cite{Domenech:2024wao,Papanikolaou:2020qtd,Domenech:2023jve,Domenech:2020ssp}
\begin{equation}\label{eq:mass-range}
    10\text g\lesssim M_{\text{PBH},0}\lesssim10^{9}\text g.
\end{equation}
Since the position of the Gaussian peak sets the mean mass of the distribution of PBH, we choose $k_\text{peak}$ to fit accordingly inside this mass range. Although the parametrization of the peak of the power spectrum is rather arbitrary, we, in essence, try to reproduce a Starobinsky-like model \cite{Starobinsky:1980te} with a feature in the potential that produces the peak, as explained in the introduction. This allows us to select the values of the decay rates in terms of the expected temperature of reheating of these models \cite{Bezrukov:2011gp,Gorbunov:2010bn,Jeong:2023zrv}. {Thus, we consider the following range of inflaton decay rates for our analysis}
\begin{equation}
    \Gamma_\phi^{\text{min}}=10^{-25}\Mpl<\Gamma_\phi<10^{-19}\Mpl=\Gamma_\phi^{\text{max}},
\end{equation}
which implies {reheating} temperatures of $10^6$GeV to $10^9$GeV. We have checked that $\Gamma_\phi^{\text{max}}$ does not conflict with the PBH. That is, they form before the field decays. Also, for $\Gamma_\phi^\text{min}$, we select the scenarios where PBH dominate before their evaporation (otherwise they do not produce GWs). In any case, the decay rate of the inflaton should be considered carefully. It could be the case that when the PBH dominate, the temperature of the surrounding radiation \eqref{eq:T-decay}, due to the field's decay, is higher than the temperature of the PBH themselves \eqref{eq:T-evap}, which delays the evaporation process and allows a longer PBH-dominated phase. We consider this when computing the GWs from the PBH-dominated phase in Sec.~\ref{sec:IGWs}. However, this effect on the production of GWs, if any, is minimal.

%%%%%%%%%%%%%%%%%%%%%%%%%%%%%%%%%%%
%% PRIMORDIAL BLACK HOLE DOMINANCE
%%%%%%%%%%%%%%%%%%%%%%%%%%%%%%%%%%%

\section{Primordial black hole dominance}\label{sec:PBH-dominated}

In this section, we describe the collapse process of a perturbation into a PBH under the KP formalism and compute the fractional energy density $\Omega_{\text{PBH}}$. Then, assuming that the scalar field decays into radiation, we study the evolution of the energy densities of the field, radiation, and the PBH. Finally, after determining the conditions for a PBH-dominated era, we show the power spectrum of Poissonian fluctuations of the energy density of the PBH fluid.

%%%%%%%%%%%%%%%%%%%%%%%%%%%%%%%%%%%%%

\subsection{Khlopov-Polnarev formalism}

\begin{figure}
     \centering
     \begin{subfigure}[b]{0.49\textwidth}
         \centering         \includegraphics[width=\textwidth]{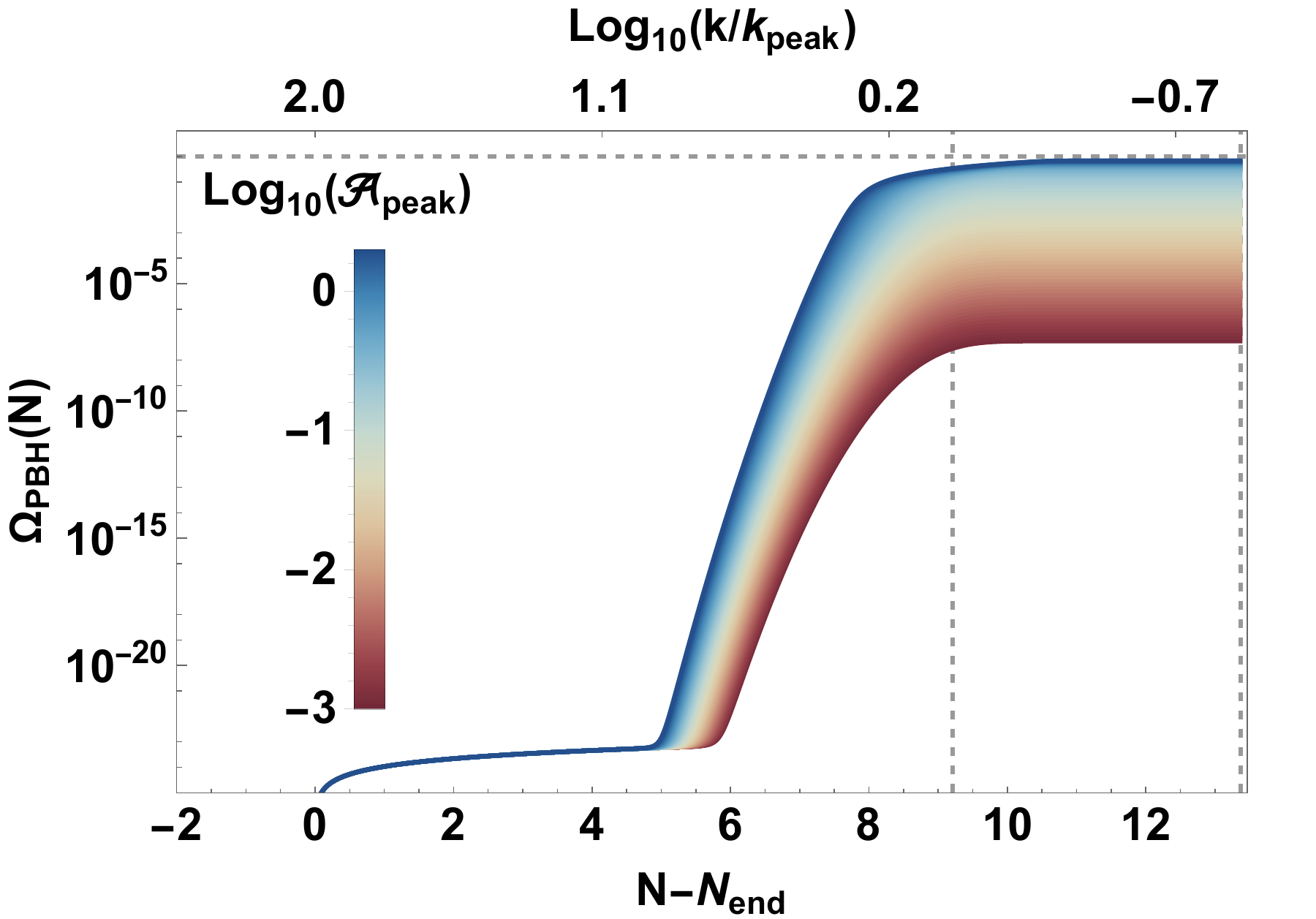}
         \caption{}
         \label{fig:Omega1}
     \end{subfigure}
     \begin{subfigure}[b]{0.49\textwidth}
         \centering         \includegraphics[width=\textwidth]{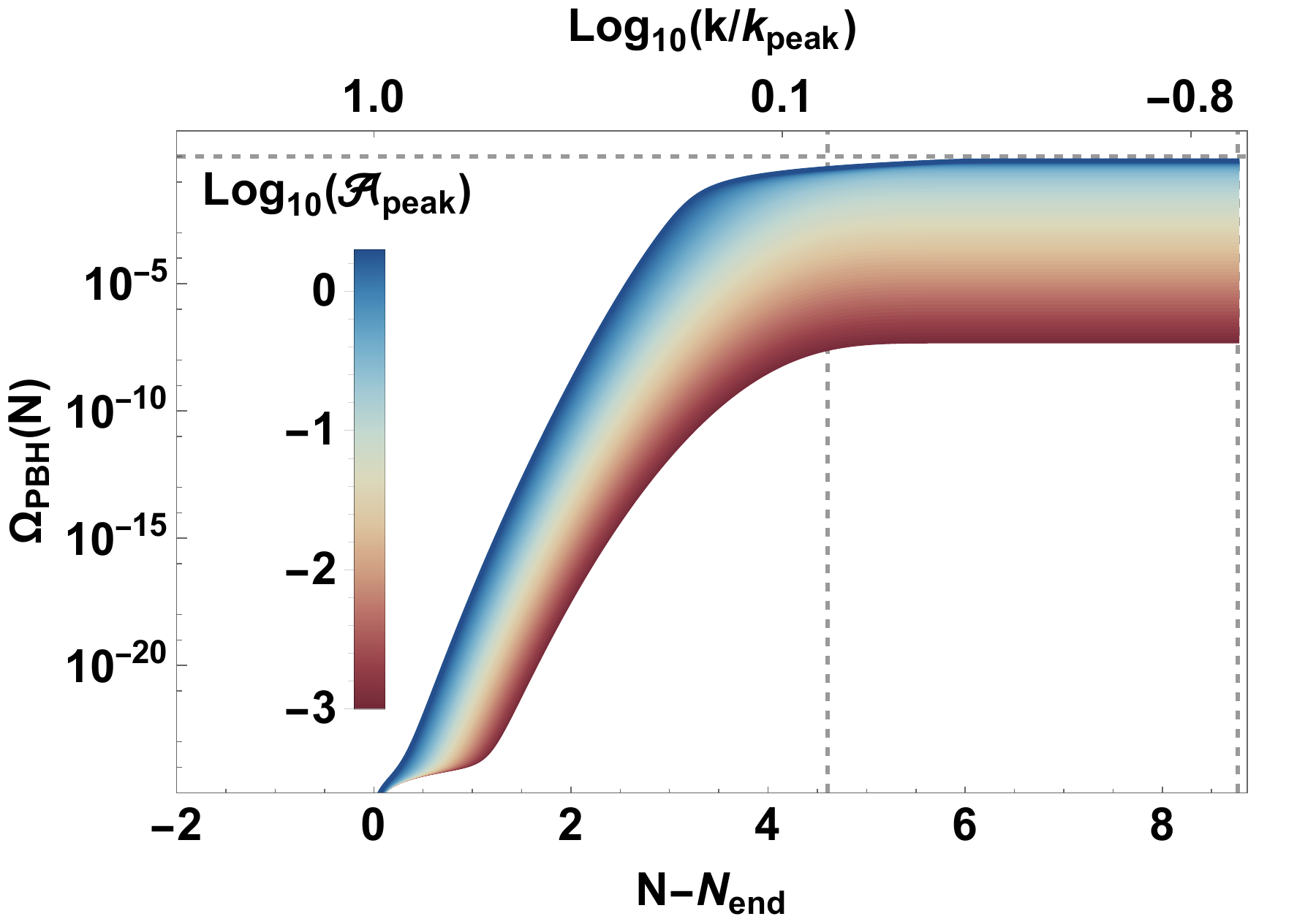}
         \caption{}
         \label{fig:Omega2}
     \end{subfigure}
        \caption{Fractional energy density of PBH for \textbf{a)} $k_{\text{peak}}=10^{-2}k_{\text{end}}$ and \textbf{b)} $k_{\text{peak}}=10^{-1}k_{\text{end}}$ as a function of the number of efolds $N$. Here, $k_{end} = 5.7\times 10^{24} {\rm Mpc}^{-1}$ is the scale corresponding to the end of inflation, which exits the horizon approximately $N_{end}=60$ e-folds after the pivot scale. 
        The vertical dashed line to the left corresponds to the point where the peak enters the horizon, and the one to the right when the scale $k_{\text{peak}}/8$ enters the horizon. The horizontal dashed line corresponds to $\Omega_{\text{PBH}}(t)=1$.}
        \label{fig:Omega}
\end{figure}

{As stated in the introduction, in a matter-dominated era, the formation probability of a PBH relies on the fraction of sufficiently spherical regions to undergo collapse. This is the initial scenario proposed by Khlopov and Polnarev in the 80's \cite{Khlopov:1980mg,Khlopov:1982ef,Polnarev:1985btg,Khlopov:2008qy,Polnarev:1981}. In an almost spherical collapse, gravity pulls matter radially inward toward the center, but in an anisotropic collapse, matter collapses faster in some directions than in others. If these differences are significant, shear stresses can disrupt the formation of a PBH \cite{Barrow:1978}. However, a moderate anisotropy can allow collapse. For instance, if a perturbation is slightly elongated or deformed but still retains a strong central gravitational potential, it can collapse into a PBH. This is computed by taking into account the Zel’dovich approximation for the nonlinear evolution of density perturbations, Thorne’s hoop conjecture, and the probability distribution for nonspherical perturbations derived by Doroshkevich; see \cite{Harada:2016mhb} for details. The original analysis \cite{Khlopov:1980mg} gave $\beta(k)\simeq0.02\sigma_k^5$, which was also later refined in \cite{Harada:2016mhb} to obtain the semi-analytical formula $\beta(k)\simeq0.056\sigma_k^5$, which is valid for $\sigma_k\lesssim10^{-2}$. Two important remarks are worth mentioning here (1) These estimations for $\beta(k)$ are based on analytical fits of full numerical computations under the assumption of small perturbations, so in this sense, they do not capture the physics of scenarios with amplified perturbations. Thus, since in this work, we consider amplified perturbations, we have derived in App.~\ref{sec:appendix}, and for the first time, an improved semi-analytical formula for $\beta(k)$ valid for $\sigma_k$ up to $\mathcal{O}(1)$ that recovers the previous one for $\sigma_k\lesssim10^{-2}$. This is given by:
\begin{equation}\label{eq:KP-out-of-appendix}
    \beta(k)=\frac{A_1\,\sigma_k^5+A_2\,\sigma_k^6}{1+A_1\,\sigma_k^5+A_3\,\sigma_k^6},
\end{equation}
where the constants $A_1$, $A_2$, and $A_3$ are defined in App.~\ref{sec:appendix}. This is the formula for the mass fraction that we are using in our computations. (2)} the mass fraction must be computed, for each $k$, at the time $t_k$ when the mode enters the horizon. This is obtained by assuming that when a mode crosses the horizon the relation $k=a(t_k)H(t_k)$ is satisfied, which during a matter-dominated phase is given in terms of $k$ by
\begin{equation}\label{eq:hubble-crossing-time}
    t_k\simeq t_{\text{end}}\left(\frac{k_{\text{end}}}{k}\right)^3,
\end{equation}
where we have used eqns.~\eqref{eq:scale-factor} and \eqref{eq:hubble-rate}, and defined $k_{\text{end}}=a(t_{\text{end}})H(t_{\text{end}})$. Thus, using \eqref{eq:KP-out-of-appendix}, we can compute the mass fraction as a function of time and then relate it to the fractional energy density in the form of PBH as follows \cite{Press:1973iz,Harada:2013epa}
\begin{equation}\label{eq:Omega-PBH}
    \beta(M_\text{f})\equiv\frac{\dd\Omega_{\text{PBH}}(M_\text{f})}{\dd\ln(M_{\text{f}})}\qquad\rightarrow\qquad\Omega_{\text{PBH}}(M_\text{f})=\int_{M_\text{H}^\text{end}}^{M_\text{f}}\beta(\tilde M_\text{f})\dd\ln(\tilde M_{\text{f}}),
\end{equation}
where $M_\text{f}$ is the PBH mass at the moment of formation and the lower limit of the integral is the horizon mass at the end of inflation, which corresponds to the smallest possible PBH mass. As explained above, to not to overestimate $\Omega_\text{PBH}$, we must take into account the time at which each perturbation enters the horizon and collapses. To do this, we relate the mass of the PBH at formation, $M_\text{f}$, with the wavenumber $k$ at the moment it crosses the horizon using equation \eqref{eq:hubble-crossing-time} as follows
\begin{equation}\label{eq:PBH-mass}
    M_{\text{f}}(k)\simeq\frac{4\pi\gamma}{H(t_k)}\simeq\gamma M_H^{\text{end}}\left(\frac{k_{\text{end}}}{k}\right)^3,
\end{equation}
where $\gamma$ determines the fraction of the horizon mass that goes into the PBH (we assume $\gamma=1$ for simplicity) and $M_H^{\text{end}}$ is the horizon mass at the end of inflation. Using \eqref{eq:PBH-mass}, the wavenumber $k$ can be seen as a ``measure of time'' and the integral in \eqref{eq:Omega-PBH} is rewritten as
\begin{equation}\label{eq:Omega-PBH-k}
    \Omega_\text{PBH}(k)=3\int_k^{k_\text{end}}\beta(\tilde k)\dd\ln(\tilde k),
\end{equation}
where the lower limit represents the moment in cosmic time $t$ at which the wavenumber $k$ crosses the horizon, \textit{c.f.} eqn.~\eqref{eq:hubble-crossing-time}, and the factor of 3 comes from the relation between $\dd\ln(M_f)$ and $\dd\ln(k)$, see eqn.~\eqref{eq:PBH-mass}. The results for the numerical integration of \eqref{eq:Omega-PBH-k} are shown in Fig.~\ref{fig:Omega} as a function of the number of efolds $N$ from the end of inflation ($N_{\text{end}}$) for two values of the position of the Gaussian peak, $10^{-2}k_{\text{end}}$, and $10^{-1}k_{\text{end}}$, with its amplitude ranging from $10^{-3}$ to $1$. We observe that as we move the Gaussian peak towards higher scales, it takes more time for $\Omega_{\text{PBH}}$ to grow and (potentially) dominate. Moreover, we also observe the appearance of a plateau for high values of $N$, whose amplitude depends on $\mathcal{A}_{\text{peak}}$, see App.~\ref{sec:appendix2} for details. This clearly reflects the fact that the inflationary spectrum $\mathcal{P}_{\mathcal{R}}^{\text{inf}}$ hardly contributes to the mass fraction, and once the peak has entered the horizon, $\Omega_{\text{PBH}}$ stays fixed to a constant value. {We consider that the totality of the peak has entered the horizon when the scale $k_\text{peak}/8$ enters, which approximately corresponds to a distance of $10\sigma$ from the peak.} This is the reason why the numerical computation of \eqref{eq:Omega-PBH} is terminated at the moment $t_{k_\text{peak}/8}$, represented by the vertical dashed line to the right, when the scale $k_{\text{peak}}/8$ enters the horizon and $\Omega_{\text{PBH}}$ is fixed on the plateau. However, we remark that the choice of $10\sigma$ is just orientative, and one could, in principle, choose either higher or lower values for this scale. In App.~\ref{sec:appendix2}, we show some analytical approximations to the fractional energy density of PBH \eqref{eq:Omega-PBH-k}, where the dependence with the parameters of the model, specifically with $\mathcal A_\text{peak}$, is revealed.

{In principle, one could consider other non-spherical effects in the formation probability of the PBH under the KP formalism, such as the inhomogeneity effect \cite{Polnarev:1981,Kokubu:2018fxy,Harada:2016mhb}, or the angular momentum of the black holes \cite{Harada:2017fjm}. If that is the case, the total mass fraction is computed as the product of the individual mass fractions associated with each effect. We however restrict this work to the anisotropy effect, as it is the dominant one.}

%%%%%%%%%%%%%%%%%%%%%%%%%%%%%%%%%%%%%

\subsection{Hawking evaporation}

As seen in Fig.~\ref{fig:Omega}, once the Gaussian peak has entered the horizon, the fractional energy density of PBH reaches the plateau and gets fixed to a constant value. For reasons obvious in the following, we call this time the ``initial'' time and label it as 
\begin{equation}\label{eq:initial-time}
    t_{\text{in}}=t_{k_\text{peak}/8}.
\end{equation}
According to Hawking \cite{Hawking:1974rv}, PBH evaporates and emits particles with an approximate thermal spectrum corresponding to the temperature $T_{\text{PBH}}=\Mpl^2/M_{\text{PBH}}$ (based on Hawking evaporation). However, due to this particle emission, the PBH loses mass at a rate given by \cite{Page:1976df,Hooper:2019gtx,Domenech:2024wao}
\begin{equation}\label{eq:decay-rate}
    \Gamma_{\text{PBH}}(t)\equiv-\frac{\dd\ln M_{\text{PBH}}(t)}{\dd t}=A\frac{\Mpl^4}{M_{\text{PBH}}^3(t)},
\end{equation}
where the constant $A$ is given by
\begin{equation}
    A=\frac{3.8\pi}{480}g_H,
\end{equation}
where we assume $g_H\simeq108$ (corresponding to standard model degrees of freedom) for $M_{\text{PBH}}\ll10^{11}g$. In our case, we have an extended mass distribution, contrary to the standard monochromatic case. To simplify, we assume that the mean PBH mass corresponds to the mass of the PBH formed at the peak of the distribution, that is
\begin{equation}\label{eq:mean-PBH-mass}
    M_{\text{PBH},0}=M_\text{f}(k_\text{peak}).
\end{equation}
By integrating eqn.~\eqref{eq:decay-rate}, one obtains the following time-dependence of the mass
\begin{equation}
    M_{\text{PBH}}(t)=M_{\text{PBH},0}\left(1-\frac{t}{t_{\text{eva}}}\right)^{1/3},
\end{equation}
being $t_{\text{eva}}$ the time at which PBH completely evaporates, given by
\begin{equation}\label{eq:evaporation-time}
    t_\text{eva}=\frac{M_{\text{PBH},0}^3}{3A\Mpl^4}.
\end{equation}

%%%%%%%%%%%%%%%%%%%%%%%%%%%%%%%%%%%%%%

\subsection{Boltzmann equations for the evolution of the energy densities}

Considering now the energy transfer from the inflaton field to the PBH and then to radiation, we consider the following Boltzmann equations for the evolution of the energy densities \cite{Domenech:2024wao}:
\begin{equation}\label{eq:Boltzmann-equations}
\begin{split}
    \dot{\rho}_{\text{PBH}}+(3H+\Gamma_\text{PBH})\,\rho_\text{PBH}&=0\\
    \dot{\rho}_{\phi}+(3H+\Gamma_\phi)\,\rho_\phi&=0\\
    \dot{\rho}_{\text{r}}+4H\rho_\text{r}-\Gamma_\text{PBH}\,\rho_\text{PBH}-\Gamma_\phi\,\rho_\phi&=0,
\end{split}
\end{equation}
where $\rho_\text{PBH}$, $\rho_\phi$, and $\rho_\text{r}$ represent the energy densities of PBH, scalar field, and radiation, respectively. To solve the system \eqref{eq:Boltzmann-equations} we use the following set of initial conditions
\begin{equation}\label{eq:initial-conditions}
\begin{split}
    \rho_\text{PBH}(t_\text{in})&=\Omega_\text{PBH}(t_\text{in})\rho_T(t_\text{in})\\
    \rho_\phi(t_\text{in})&=(1-\Omega_\text{PBH}(t_\text{in}))\rho_T(t_\text{in})\\
    \rho_\text{r}(t_\text{in})&=0,
\end{split}
\end{equation}
where the total energy density, $\rho_T$, follows from the Friedmann's equation
\begin{equation}
    \rho_T=\rho_\text{PBH}+\rho_\phi+\rho_\text{r}=3H^2\Mpl^2.
\end{equation}
\begin{figure}
     \centering
     \begin{subfigure}[b]{0.49\textwidth}
         \centering         \includegraphics[width=\textwidth]{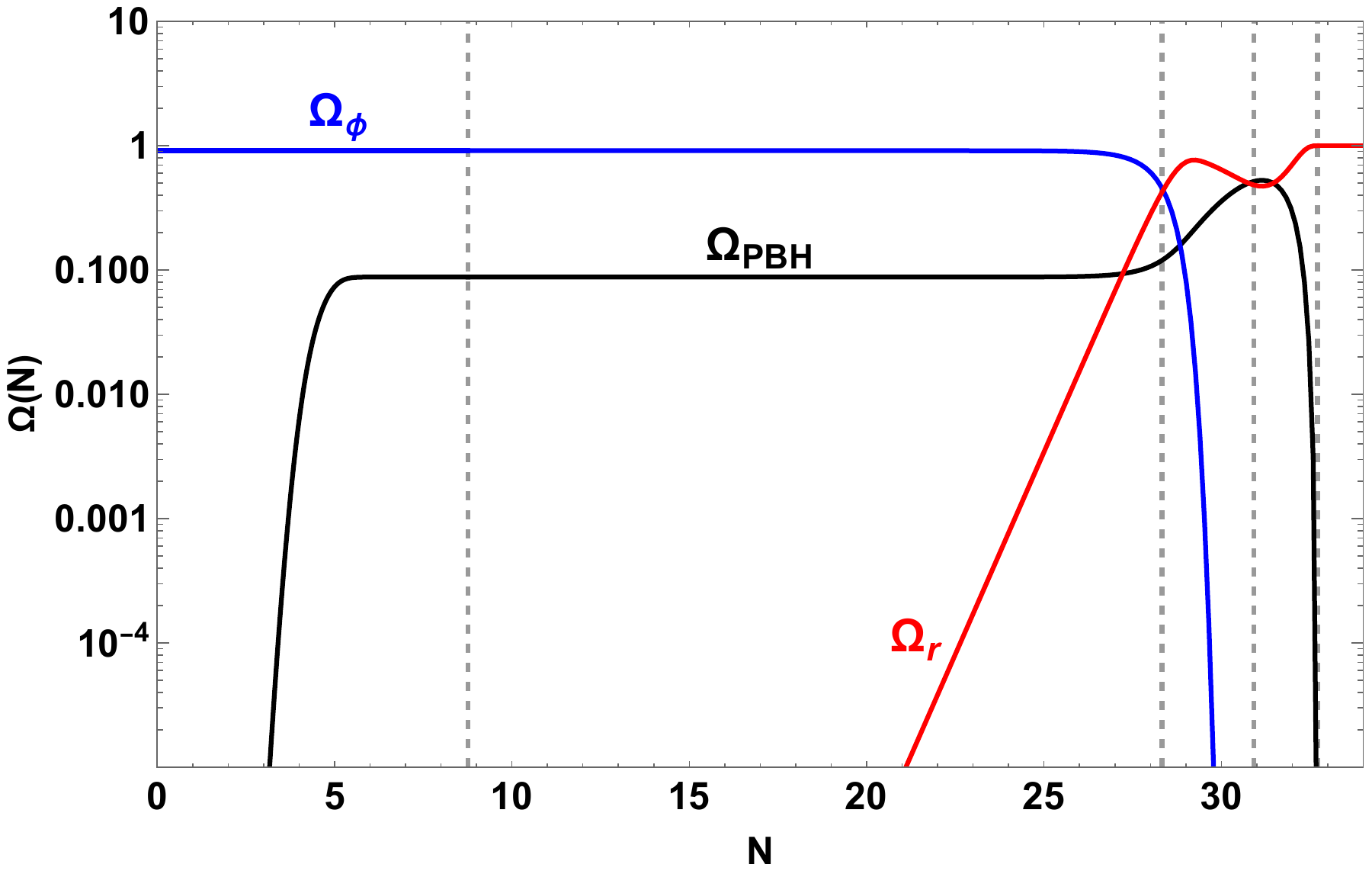}
         \caption{$\Omega_{\text{PBH}}(t_{\text{in}})=0.088$, $\Gamma_\phi=10^{-23}\Mpl$}
         \label{fig:Boltzmann11}
     \end{subfigure}
     \begin{subfigure}[b]{0.49\textwidth}
         \centering         \includegraphics[width=\textwidth]{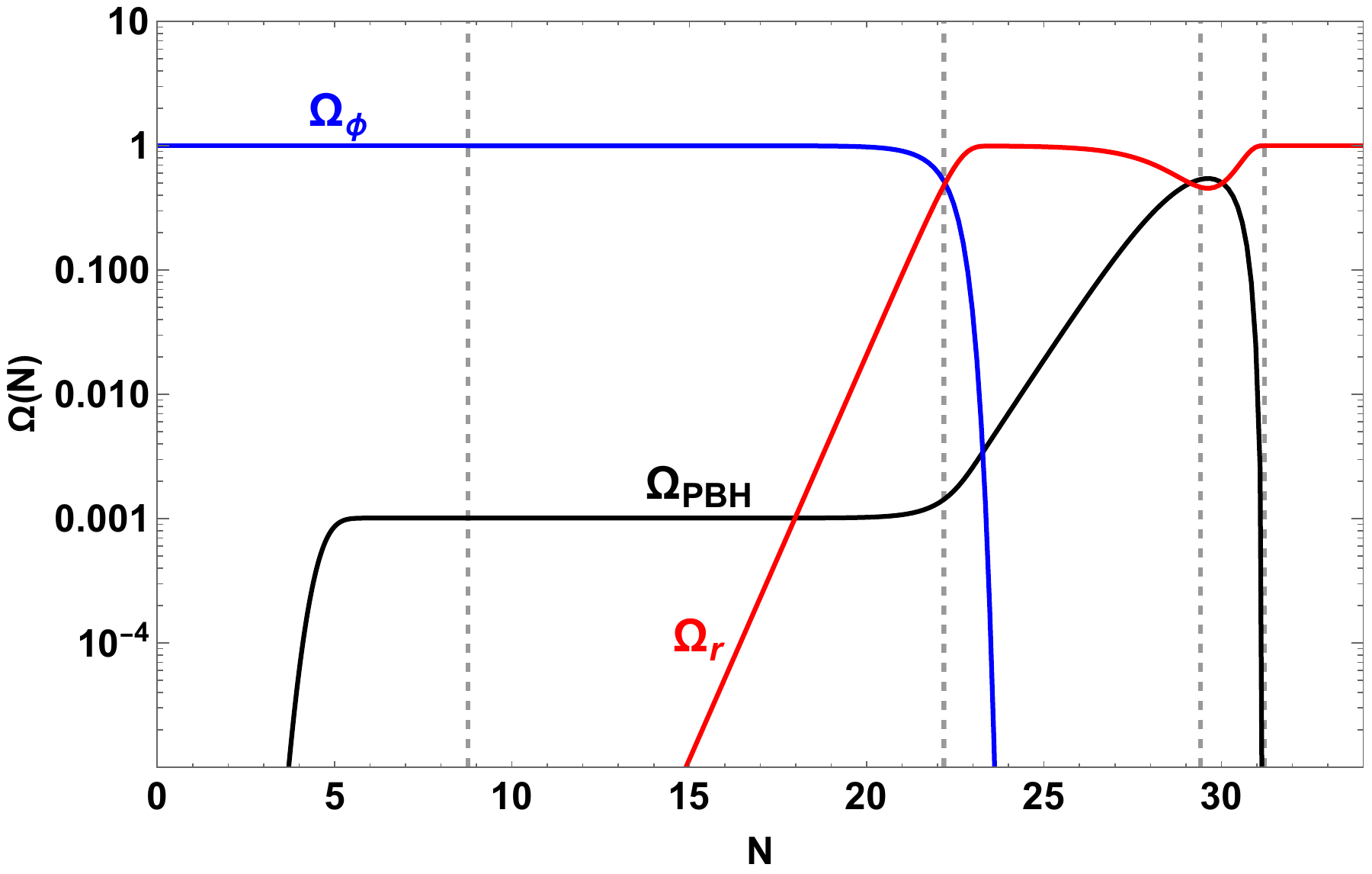}
         \caption{$\Omega_{\text{PBH}}(t_{\text{in}})=0.001$, $\Gamma_\phi=10^{-19}\Mpl$}
         \label{fig:Boltzmann21}
     \end{subfigure}
     \begin{subfigure}[b]{0.49\textwidth}
         \centering         \includegraphics[width=\textwidth]{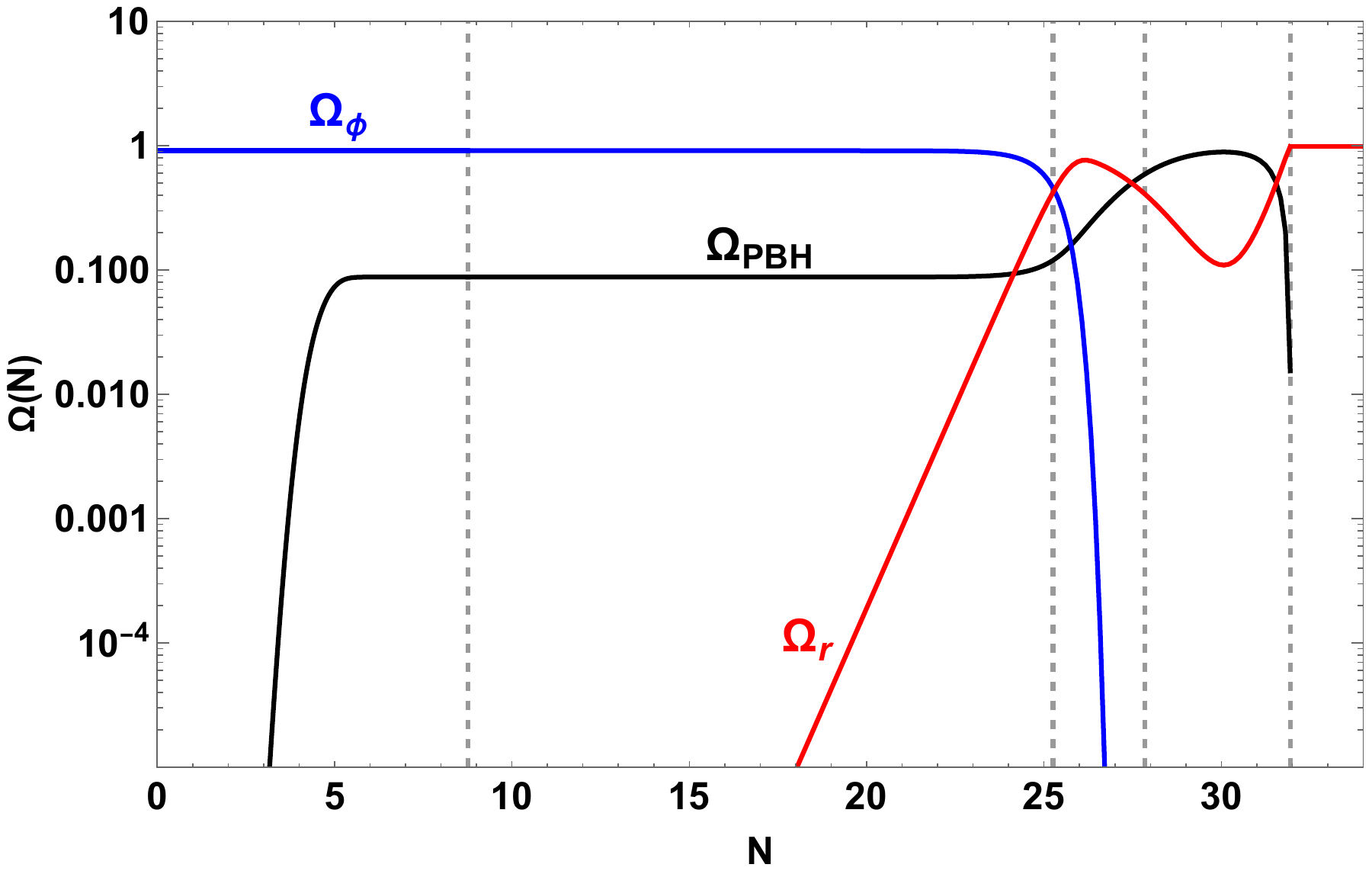}
         \caption{$\Omega_{\text{PBH}}(t_{\text{in}})=0.088$, $\Gamma_\phi=10^{-21}\Mpl$}
         \label{fig:Boltzmann12}
     \end{subfigure}
     \begin{subfigure}[b]{0.49\textwidth}
         \centering         \includegraphics[width=\textwidth]{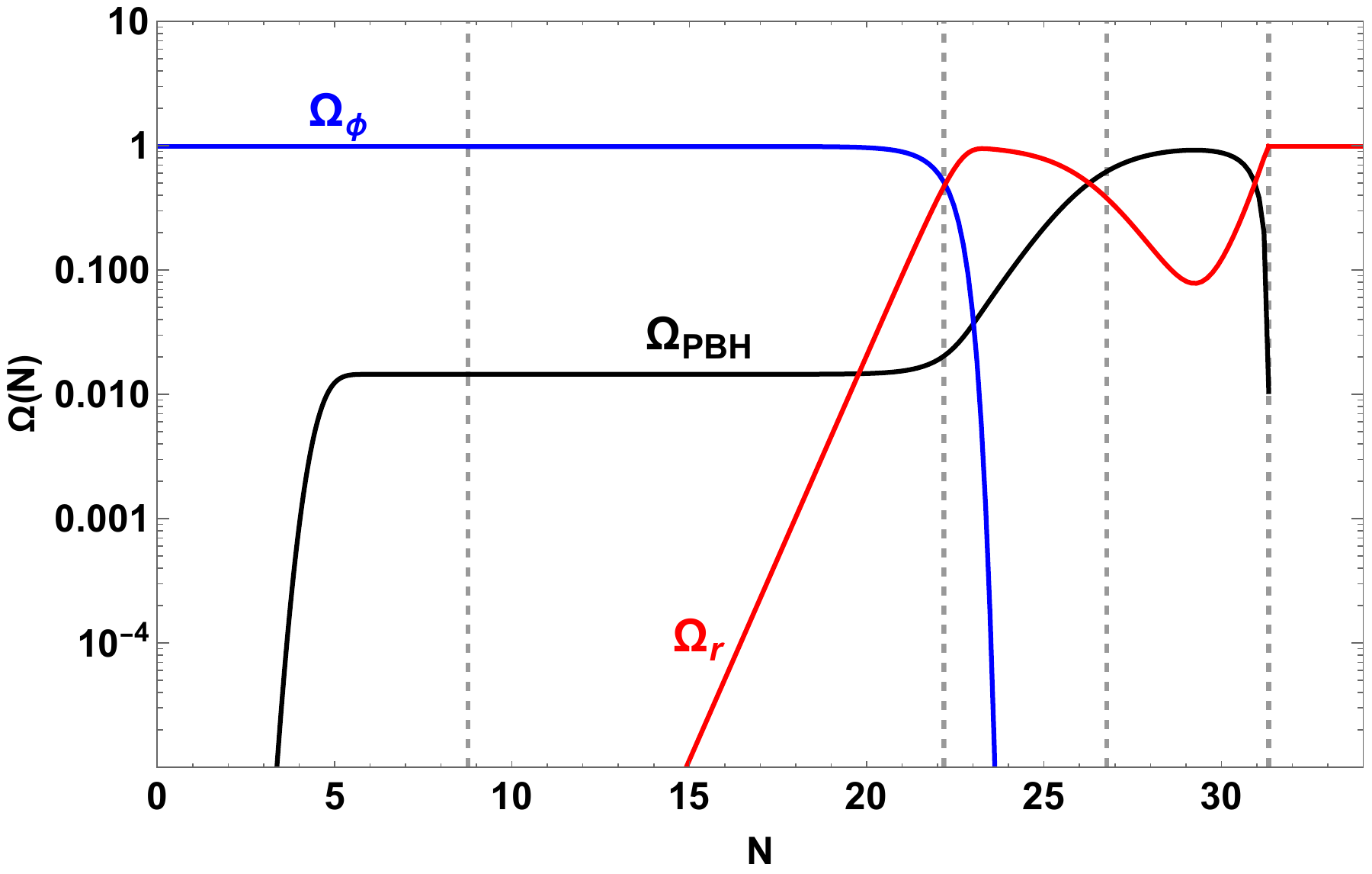}
         \caption{$\Omega_{\text{PBH}}(t_{\text{in}})=0.015$, $\Gamma_\phi=10^{-19}\Mpl$}
         \label{fig:Boltzmann22}
     \end{subfigure}
     \begin{subfigure}[b]{0.49\textwidth}
         \centering         \includegraphics[width=\textwidth]{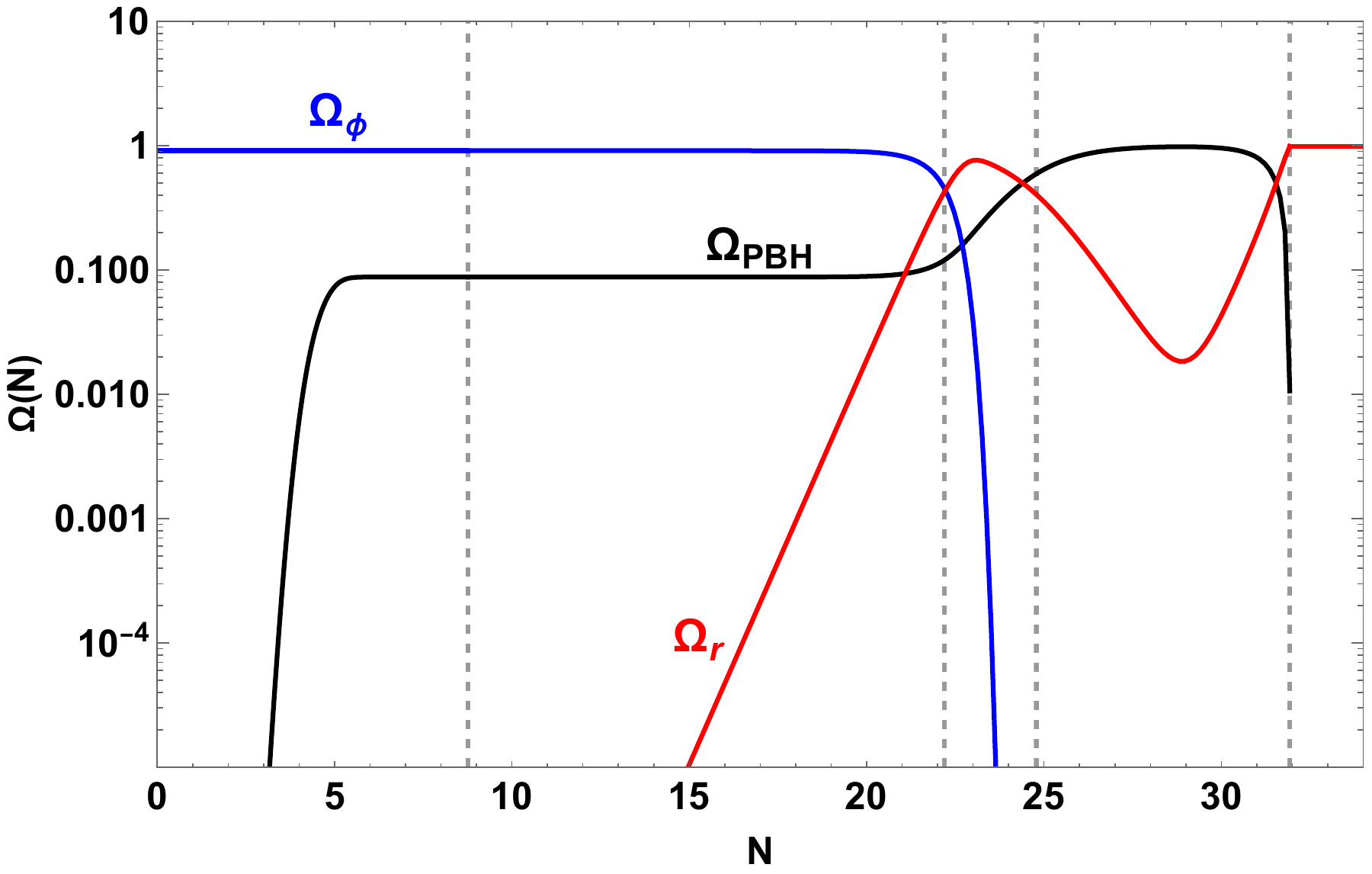}
         \caption{$\Omega_{\text{PBH}}(t_{\text{in}})=0.088$, $\Gamma_\phi=10^{-19}\Mpl$}
         \label{fig:Boltzmann13}
     \end{subfigure}
     \begin{subfigure}[b]{0.49\textwidth}
         \centering         \includegraphics[width=\textwidth]{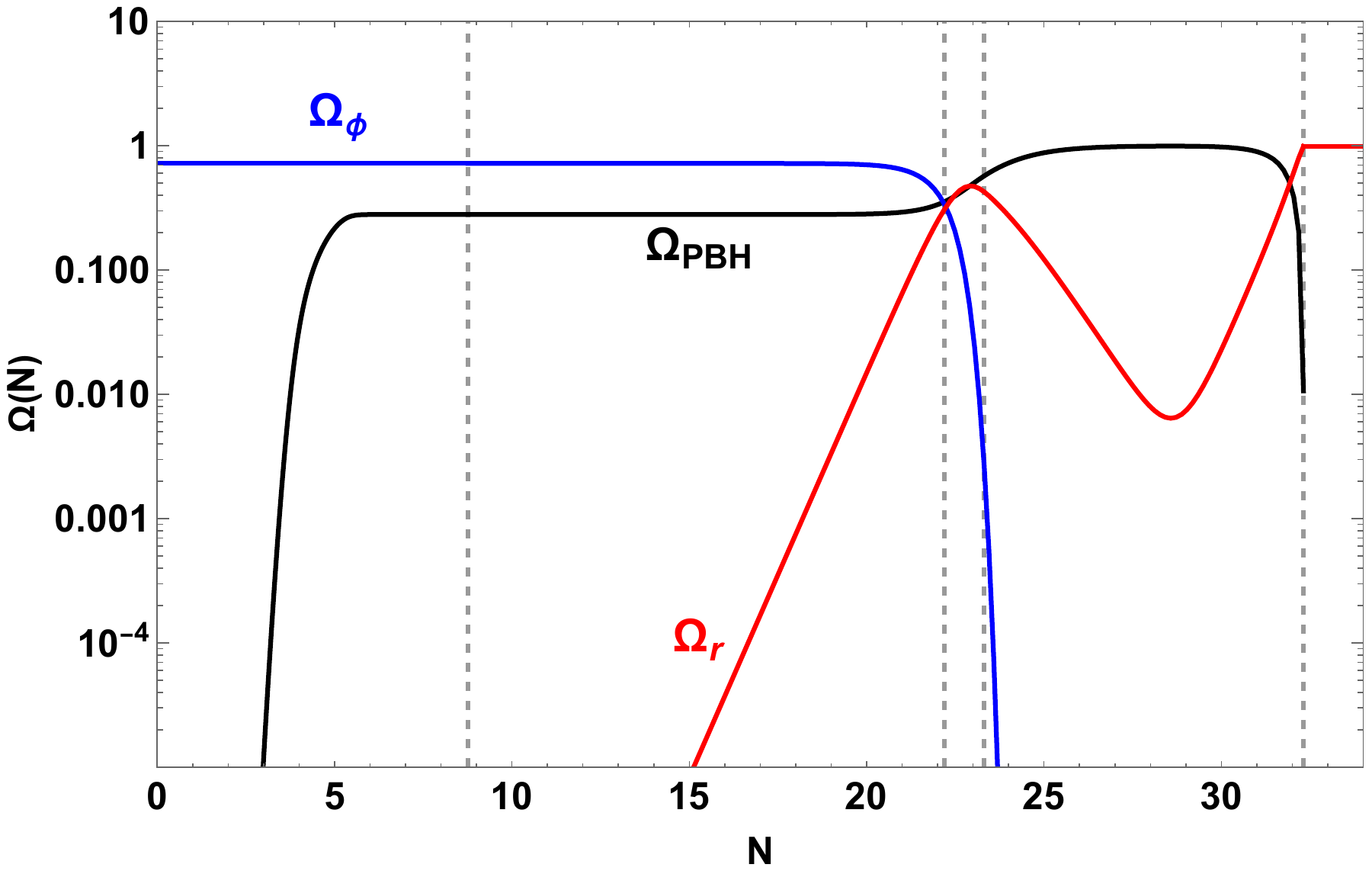}
         \caption{$\Omega_{\text{PBH}}(t_{\text{in}})=0.230$, $\Gamma_\phi=10^{-19}\Mpl$}
         \label{fig:Boltzmann23}
     \end{subfigure}
        \caption{Evolution of the fractional energy densities for PBH (black), the scalar field (blue), and radiation (red). The left column shows the effect of varying $\Gamma_\phi$, growing from top to bottom, whereas the right column shows the effect of a varying $\Omega_{\text{PBH}}(t_\text{in})$, also growing from top to bottom. The vertical dashed lines marks from left to right: $t_{\text{in}}$ (translated to efolds), $\phi$-radiation equality, radiation-PBH equality, and $t_\text{eva}$ (translated to efolds), respectively. In Fig.~\ref{fig:Boltzmann13} $t_\text{in}$ and $\phi$-radiation equality coincide.}
        \label{fig:Boltzmann}
\end{figure}
Fig.~\ref{fig:Boltzmann} shows the solution of the system \eqref{eq:Boltzmann-equations} together with the initial conditions \eqref{eq:initial-conditions} and for several values of $\Gamma_\phi$ and $\Omega_{\text{PBH}}(t_\text{in})$. {Note that the values of $\Omega_{\text{PBH}}(t_{\rm in})$ in our case are not arbitrary, rather they are determined from the PBH formation mechanism which in our case is determined by KP framework we explained in the previous section. The time $t_{\rm in}$ here can also be seen as the time when the PBH critical energy density $\Omega_{\rm {PBH}}$ gets saturated to a constant value (See Fig.~\ref{fig:Omega}). }
We have chosen $k_\text{peak}=0.1k_\text{end}$ for all our computations. The left column in Fig.~\ref{fig:Boltzmann} shows the effect of changing $\Gamma_\phi$, growing from top to bottom, whereas the right column shows the effect of a varying $\Omega_{\text{PBH}}(t_\text{in})$, also growing from top to bottom. We observe that the higher both of these parameters are, the longer the PBH dominates. Since the position of the Gaussian peak is the same for all plots, which essentially sets the mean mass of the PBH distribution, evaporation occurs (approximately) at the same number of efolds, $N_\text{eva}$\footnote{{It is worth noting here that any} modification of $\Gamma_\phi$ and/or $\mathcal A_{\text{peaks}}$ turns into different expansion rates of the universe, depending on the amount of time matter ($\phi$, PBH) or radiation dominates. {This means even though $t_{\rm eva}$ is fixed $N_{\rm eva}$ could in principle differ based on the duration of PBH dominance (See \eqref{eq:delta-PBH}). }}. Thus, setting $k_\text{peak}$ towards higher scales increases the evaporation time and consequently decreases the temperature of the plasma after evaporation, which could conflict with BBN, as shown above (see eqn.~\eqref{eq:mass-range}).

The vertical dashed lines in Fig.~\ref{fig:Boltzmann} mark several relevant times, translated to number of efolds. From left to right: $t_{\text{in}}$, $\phi$-radiation equality ($t_\text{r}$), radiation-PBH equality ($t_\text{PBH}$), and PBH evaporation ($t_\text{eva}$). The former and the latter are defined in eqns.~\eqref{eq:initial-time} and \eqref{eq:evaporation-time}, respectively. The $\phi$-radiation equality occurs approximately at $t_\text{r}\sim\Gamma_\phi^{-1}$ and, to estimate the transition from radiation to PBH we do the following. Initially, the scalar field dominates, and thus $\rho_\phi$ redshifts as
\begin{equation}
    \rho_\phi(t)\sim\rho_\phi(t_\text{in})\left(\frac{a(t_\text{in})}{a(t)}\right)^3\sim\rho_\phi(t_\text{in})\left(\frac{t_\text{in}}{t}\right)^2,
\end{equation}
where we assume the standard matter-dominated power-law behavior of the scale factor, $a\sim t^{2/3}$. The decay into radiation occurs at a time $t_\text{r}\sim\Gamma_\phi^{-1}$, which implies that
\begin{equation}
    \rho_\phi(t_\text{r})\sim\rho_\phi(t_\text{in})\left(\frac{t_\text{in}}{t_\text{r}}\right)^2.
\end{equation}
At this point, $\rho_\phi(t_\text{r})\sim\rho_r(t_\text{r})$ and radiation becomes the dominant component. This implies that $\rho_\text{r}$ redshifts as
\begin{equation}\label{eq:r-redshift}
    \rho_\text{r}(t)\sim\rho_\text{r}(t_\text{r})\left(\frac{a(t_\text{r})}{a(t)}\right)^4\sim\rho_\text{r}(t_\text{r})\left(\frac{t_\text{r}}{t}\right)^2,
\end{equation}
where now we used that $a\sim t^{1/2}$ during radiation-dominated. On the other side, $\rho_\text{PBH}$ redshifts as follows during this radiation-dominated era
\begin{equation}\label{eq:PBH-redshift}
    \rho_\text{PBH}(t)\sim\rho_\text{PBH}(t_\text{in})\left(\frac{t_\text{in}}{t_\text{r}}\right)^2\left(\frac{t_\text{r}}{t}\right)^{3/2},
\end{equation}
where we considered the redshift during the time $\rho_\phi$ dominates. Equating \eqref{eq:r-redshift} and \eqref{eq:PBH-redshift} we obtain, approximately, the time at which PBH dominate, which is given by
\begin{equation}\label{eq:PBH-time}
    t_\text{PBH}\sim\Gamma_\phi^{-1}\left(\frac{1-\Omega_\text{PBH}(t_\text{in})}{\Omega_\text{PBH}(t_\text{in})}\right)^2.
\end{equation}  
Using this, the approximated time the PBH-dominated phase lasts is computed as the difference between eqns~\eqref{eq:evaporation-time} and \eqref{eq:PBH-time}, that is
\begin{equation}\label{eq:delta-PBH}
    \Delta_\text{PBH}=t_\text{eva}-t_\text{PBH}=\frac{M_\text{f}^3}{3A\Mpl^4}-\Gamma_\phi^{-1}\left(\frac{1-\Omega_\text{PBH}(t_\text{in})}{\Omega_\text{PBH}(t_\text{in})}\right)^2.
\end{equation}
The longer the PBH-dominated phase lasts, the higher the induced GWs produced (See Sec.~\ref{sec:IGWs}). In this sense, eqn.~\eqref{eq:delta-PBH} clearly reflects the impact that the different parameters of the model have on the production of GWs. For instance, a small decay rate (the scalar field decays late in time), reduces the time the PBH dominate but can be compensated if their abundance is large. On the other hand, a large mean PBH mass increases this time since PBH evaporate later. In essence, eqn.~\eqref{eq:delta-PBH} shows the rich interplay between the parameters of the model.

%%%%%%%%%%%%%%%%%%%%%%%%%%%%%%%%%%%%%

\subsection{Power spectrum of primordial black hole fluctuations}

In this section, we compute the power spectrum corresponding to the PBH domination phase. In order to do this, let us assume that PBH are randomly distributed in space, meaning that the probability distribution of each PBH's position is uniform and that the locations of different black holes are uncorrelated. This assumption effectively corresponds to Poissonian statistics. A key point to note is that this description breaks down at distances smaller than the mean comoving separation, $\bar r$, between neighboring black holes. Below $\bar r$, the discrete nature of the PBH distribution becomes significant, making the fluid approximation inadequate. This mean separation is computed as \cite{Papanikolaou:2020qtd}
\begin{equation}
    \bar{r}=\left(\frac{3M_{\text{PBH},0}}{4\pi\rho_\text{PBH}}\right)^{1/3},
\end{equation}
The fact that PBH are discrete objects introduces inhomogeneities, which can be quantified with the power spectrum of PBH density fluctuations $\delta_{\text{PBH},\bm k}$. For Poissonian statistics, it is given by\footnote{{In this paper, we are not considering possible deviation from Poisson statistics by non-Gaussianities that could further lead to primordial clustering \cite{Papanikolaou:2024kjb,He:2024luf} because our consideration of the KP collapse mechanism is restricted to overdensities that are small enough. } } \cite{Papanikolaou:2020qtd}
\begin{equation}\label{eq:power-spectrum-density-PBH}
    \mathcal P_{\delta_\text{PBH}}(k)=\frac{2}{3\pi}\left(\frac{k}{k_{\text{UV}}}\right)^3\Theta\left(k_{\rm UV}-k\right),
\end{equation}
where $k_{\text{UV}}=a/\bar r$ is the ultraviolet cutoff of the power spectrum. For scales larger than $k_{\text{UV}}$, the PBH fluid can be treated as non-relativistic matter, but as $k$ approaches $k_{\text{UV}}$, the discrete nature of the PBH leads to shot-noise effects and the fluid picture is no longer valid \cite{Domenech:2023jve,Domenech:2024wao}. These energy density fluctuations correspond to isocurvature perturbations when the PBH form. In other words, PBH form in this case from the perturbations of the scalar field, and by conservation of energy, any missing part of the scalar field fluid that ends up into a PBH is compensated by PBH themselves, so that the fluctuations in both fluids are equal and opposite, \textit{i.e.}, $\delta_{\text{PBH},\bm k}=-\delta_{\phi,\bm k}$, which is what mainly characterizes isocurvature perturbations. Initially, these isocurvature PBH perturbations do not induce GWs, but as the PBH become the dominant component of the universe, the isocurvature PBH perturbations source curvature perturbations, which have a gravitational potential associated, and this last is responsible for inducing GWs. { Note that these induced GWs should not be confused with the induced GWs from amplified perturbations during, for instance, an early matter-dominated period \cite{Domenech:2021ztg,Baumann:2007,Ananda:2006,Assadullahi:2009,del-Corral:2025fzz}. In this case, the GWs are sourced by the gravitational potential of the Poissonian fluctuations associated with the overproduction (dominance) of the PBH, instead of being sourced by the inflaton perturbations. In essence, the novelty of this approach resides in the fact that the isocurvature perturbations are sourced by the scalar field, contrary to the standard approach where the radiation fluid sources the isocurvature perturbations \cite{Papanikolaou:2020qtd,Domenech:2023jve,Domenech:2024wao,Domenech:2020ssp,Papanikolaou:2022chm}. Moreover, we do take into account the whole evolution of the PBH by considering their formation mechanism, instead of assuming an initial abundance of PBH.}

So now the problem at hand is to relate the initial isocurvature fluctuations $\delta_{\text{PBH},\bm k}$ to the gravitational potential $\Phi_{\text{PBH},\bm k}$ of the PBH fluid. Following \cite{Papanikolaou:2020qtd}, this relation is given by
\begin{equation}\label{eq:grav-pot-super}
    \Phi_{\text{PBH},\bm k}\simeq -\frac{1}{5}\delta_{\text{PBH},\bm k}
\end{equation}
on super-Hubble scales, and by
\begin{equation}\label{eq:grav-pot-sub}
    \Phi_{\text{PBH},\bm k}\simeq-\frac94\left(\frac{k_\text{PBH}}{k}\right)^2\delta_{\text{PBH},\bm k}
\end{equation}
on sub-Hubble scales, where $k_\text{PBH}=a(t_\text{PBH})H(t_\text{PBH})$ is the scale that enters the horizon by the time PBH dominate, where $t_\text{PBH}$ is defined in \eqref{eq:PBH-time}. What eqns.~\eqref{eq:grav-pot-super} and \eqref{eq:grav-pot-sub} tell us is that the gravitational potential is constant in time during a PBH-dominated era, as it is expected from a matter-dominated epoch. One can interpolate between the two equations to obtain a single expression that reflects the behavior on both super and sub-Hubble scales, that is
\begin{equation}
    \Phi_{\text{PBH},\bm k}\simeq-\left[5+\frac49\left(\frac{k}{k_\text{PBH}}\right)^2\right]^{-1}\delta_{\text{PBH},\bm k},
\end{equation}
and use this in \eqref{eq:power-spectrum-density-PBH} to obtain the power spectrum of the gravitational potential associated with a dominating fluid of PBH:
\begin{equation}\label{eq:power-spectrum-density-PBH-final}
    \mathcal{P}_{\Phi_\text{PBH}}(k)=\frac2{3\pi}\left(\frac{k}{k_{\text{UV}}}\right)^3\left[5+\frac49\left(\frac{k}{k_\text{PBH}}\right)^2\right]^{-2}.
\end{equation}
This spectrum presents a maximum at $k=\frac{3\sqrt{15}}{2}k_\text{PBH}$ of order
\begin{equation}
\begin{split}
    \mathcal P_{\Phi_\text{PBH}}^{\text{max}}&=\frac{27}{64\pi}\sqrt{\frac{3}{5}}\left(\frac{k_\text{PBH}}{k_\text{UV}}\right)^3\\&\simeq\frac{27}{64\pi}\sqrt{\frac{3}{5}}\gamma\left(\frac{\Gamma_\phi}{H_\text{end}}\right)\frac{\Omega_\text{PBH}^2(t_\text{in})}{(1-\Omega_\text{PBH}(t_\text{in}))^3}\left(\frac{k_\text{end}}{k_\text{peak}}\right)^3,
\end{split}
\end{equation}
where $\Omega_\text{PBH}(t_\text{in})$ can also be estimated from the parameters of the model. See App.~\ref{sec:appendix2} and particularly eqn.~\eqref{eq:estimation-Omega} for details.

%%%%%%%%%%%%%%%%%%%%%%%%%%%%%%%%%
%% INDUCED GRAVITATIONAL WAVES
%%%%%%%%%%%%%%%%%%%%%%%%%%%%%%%%%

\section{Induced gravitational waves}\label{sec:IGWs}

Now that we have computed the power spectrum of the gravitational potential induced by the PBH density contrast, we can compute the associated GWs production due to the PBH domination \cite{Papanikolaou:2020qtd,Domenech:2023jve,Domenech:2024wao,Domenech:2020ssp,Papanikolaou:2022chm}. Before going into detail, it is important to make some remarks. GWs are mainly produced in two ways:
\begin{itemize}
    \item \textit{First-order GWs}: This signal corresponds to the stochastic background generated by the inflationary fluctuations $\mathcal P_\mathcal{R}^{\text{inf}}$, with an almost flat power spectrum and usually very weak, which we call background gravitational waves (BGWs).
    \item \textit{Second-order GWs}: Scalar perturbations couple to the tensor ones at second order in perturbation theory and induce GWs. For this reason, these are called scalar-induced gravitational waves (SIGWs). Several mechanisms can produce amplified scalar perturbations\footnote{SIGWs are also produced by the perturbations amplified during inflation \cite{Baumann:2007,Ananda:2006,Assadullahi:2009} that collapse into PBH, or even by the evaporation of these PBH \cite{Domenech:2024wao,Domenech:2020ssp}.}. However, in this work, we focus on the amplified scalar density fluctuations from the PBH-dominated phase and apply the approach in Refs.~\cite{Baumann:2007,Ananda:2006,Assadullahi:2009,Inomata:2019ivs}.
\end{itemize}
When considering both scalar and tensor fluctuations in the second-order perturbed Einstein equations, one derives the following equation of motion for the tensor perturbations
\begin{equation}\label{eq:SOEOM}
    \ddot h_{\bm k}^{(\lambda)}(t)+3H\dot h_{\bm k}^{(\lambda)}(t)+\frac{k^2}{a^2}h_{\bm k}^{(\lambda)}(t)= \frac{\mathcal{S}(\bm k,t)}{a^2},
\end{equation}
where $\lambda=+,\times$ denotes the two polarization states of the tensor modes and $\mathcal{S}(\bm k,t)$ is the source term, computed as a convolution of different modes. Here and in what follows, we work in the Newtonian gauge. The source term arises only at second order in perturbation theory and shows that the SIGWs are no longer free-propagating waves but rather a metric fluctuation arising from terms quadratic in the scalar perturbations \cite{Assadullahi:2009}. In this case, it is given by
\begin{equation}\label{eq:source-term}
    \mathcal{S}(k,t)
    =4\int\frac{\dd^3\tilde{k}}{(2\pi)^{3/2}}\tilde k^2(1-\mu^2)\Phi_{\text{PBH},\tilde{\bm k}}\Phi_{\text{PBH},|\bm k-\tilde{\bm k}|}.
\end{equation}
As one can observe from this expression, the source term reflects that the contribution from any individual mode is diluted and mixed with the contributions coming from other modes. The power spectrum of the tensor perturbations is given by
\begin{equation}\label{eq:spectrum-definition}
    \langle h_{\bm k}^{(\lambda)}(t)h_{\bm k'}^{(\lambda)}(t)\rangle=\frac12\frac{2\pi^2}{k^3}\delta^3(\bm k+\bm k')\mathcal P_h (k,t),
\end{equation}
where the $1/2$ factor comes from the fact that $\mathcal P_h(k,t)$ includes contributions from both polarizations. During a PBH-dominated phase, the gravitational potential is constant in time for all scales, and so it is the source term \eqref{eq:source-term}. Therefore, one can attempt to obtain a particular solution of \eqref{eq:SOEOM} and compute the correlator in \eqref{eq:spectrum-definition} to obtain the power spectrum of GWs induced by PBH domination:
\begin{equation}\label{eq:power-SIGWs}
    \mathcal{P}_h^{\text{PBH}}(k,t)=\frac{16g^2(k,t)}{k}\int_{k_\text{eva}}^{k_\text{UV}}\dd \tilde{k}\int_{-1}^1\dd\mu\,\frac{\tilde{k}^3\,(1-\mu^2)^2}{|k-\tilde k|^3}\mathcal P_{\Phi_\text{PBH}}(\tilde k)\mathcal P_{\Phi_\text{PBH}}(|\bm k-\tilde{ \bm k}|),
\end{equation}
where $k_{\rm eva} = a(t_{\rm eva})H(t_{\rm eva})$ that corresponds to the smallest co-moving wavenumber that enters the horizon at the PBH evaporation time. 
The function $g(k,t)$ is known as the growth function for the tensor modes, defined as
\begin{equation}
    g(k,t)=1+3\,\frac{\frac{2k}{aH}\cos \left(\frac{2k}{aH}\right)-\sin \left(\frac{2k}{aH}\right)}{\left(\frac{2k}{aH}\right)^3},
\end{equation}
see \cite{Assadullahi:2009} for details, $\mathcal P_{\Phi_\text{PBH}}(k)$ is given in \eqref{eq:power-spectrum-density-PBH-final}, and the limits of the integral are chosen so that we consider just the relevant modes during PBH-dominated. Still, the main quantity characterizing this scenario is the spectral energy density of GWs, $\Omega_{\text{GW}}^{\text{PBH}}$, used to compare theoretical predictions with current constraints and future observations. It is computed as follows
\begin{equation}\label{eq:energy-density-GWs}
    \Omega_{\text{GW}}^{\text{PBH}}(k,t)=\frac1{\rho_c}\frac{\dd \rho_{\text{GW}}}{\dd\ln{k}},
\end{equation}
and represents the energy density per logarithm of $k$ over the critical density $\rho_c=3\Mpl^2H^2$. Since GWs redshift at sub-Hubble scales as any non-interacting relativistic particles after being produced, the present spectral energy density of GWs is therefore approximated by \cite{Assadullahi:2009}
\begin{equation}\label{eq:energy-density-final}
    \Omega_{\text{GW}}^{\text{PBH}}(k,t_0)\simeq\frac{\Omega_\gamma^0}{12}\left(\frac{k}{k_{\text{eva}}}\right)^2\mathcal P_h^{\text{PBH}}(k,t_\text{eva})
\end{equation}
where $\Omega_{\gamma}^0\simeq1.2\times10^{-5}$ is the present energy density of photons, and $t_0$ represents the present epoch. See App.~\ref{sec:appendix3} for analytical estimations of \eqref{eq:energy-density-final} as a function of the parameters of the model. Particularly, eqns.~\eqref{eq:Omega-small-k} and \eqref{eq:Omega-large-k} show the approximations of $\Omega_\text{GW}^{\text{PBH}}(k,t_0)$ for $k\ll k_\text{PBH}$ and $k\gg k_\text{PBH}$, respectively. Now, these GWs are produced before BBN, and thus they cannot interfere with its predictions. If GWs are overproduced, then they contribute significantly to the radiation density and can potentially change the expansion rate of the universe, which modifies the abundance of light elements. To avoid this scenario, the total amount of energy in the form of GWs must satisfy this relation \cite{Smith:2006nka,Maggiore:1999vm}
\begin{equation}\label{eq:BBN}
    \mathcal I_\text{GW}=\int_{0}^{\infty}\Omega_{GW}^{\text{PBH}}(k)\,\dd\ln(k)\leq 1.6\times10^{-5}=\mathcal I_{\text{BBN}}.
\end{equation}
\begin{figure}
     \centering
     \begin{subfigure}[b]{0.49\textwidth}
         \centering         \includegraphics[width=\textwidth]{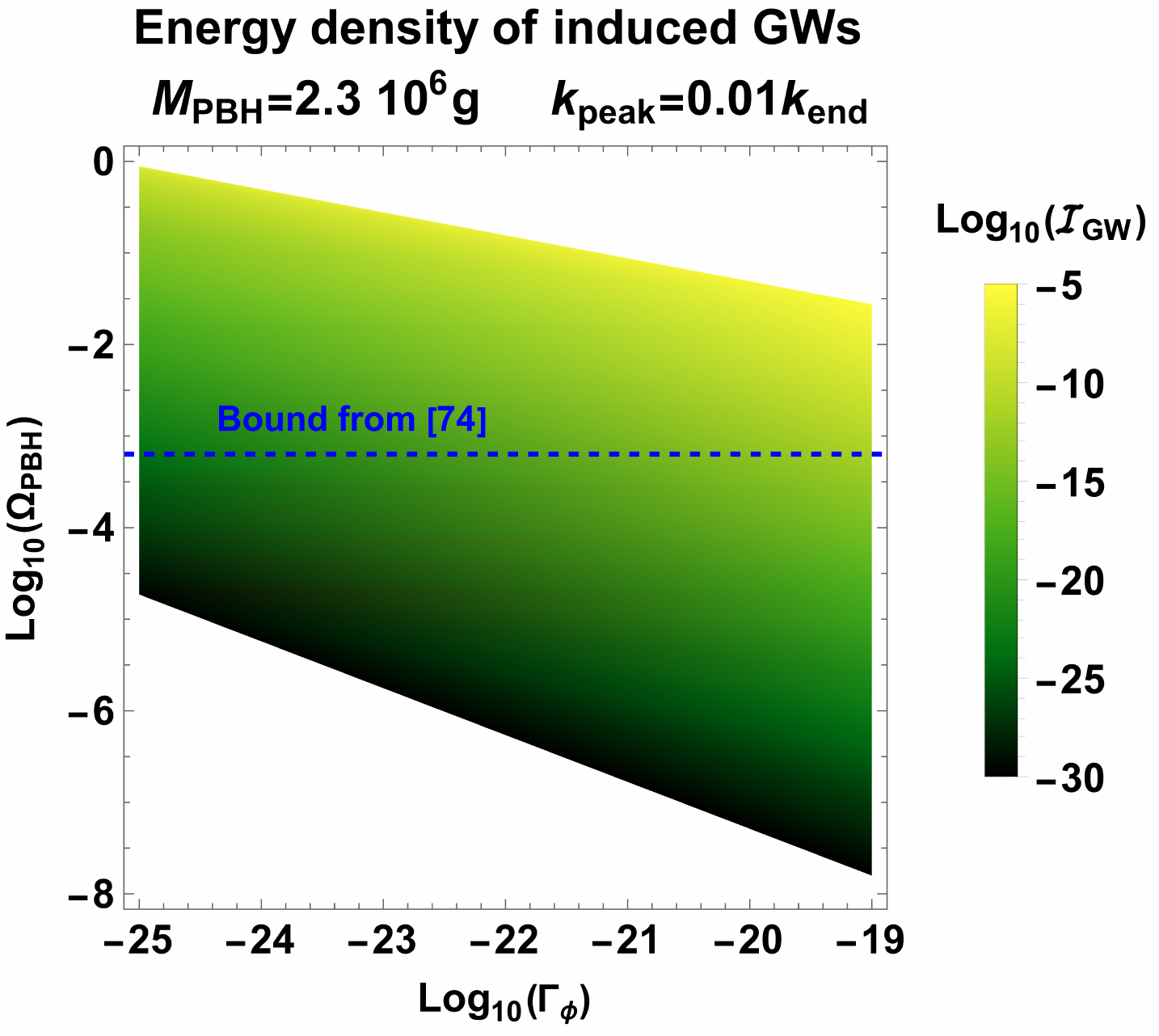}
         \caption{}
         \label{fig:GWs2}
     \end{subfigure}
     \begin{subfigure}[b]{0.49\textwidth}
         \centering         \includegraphics[width=\textwidth]{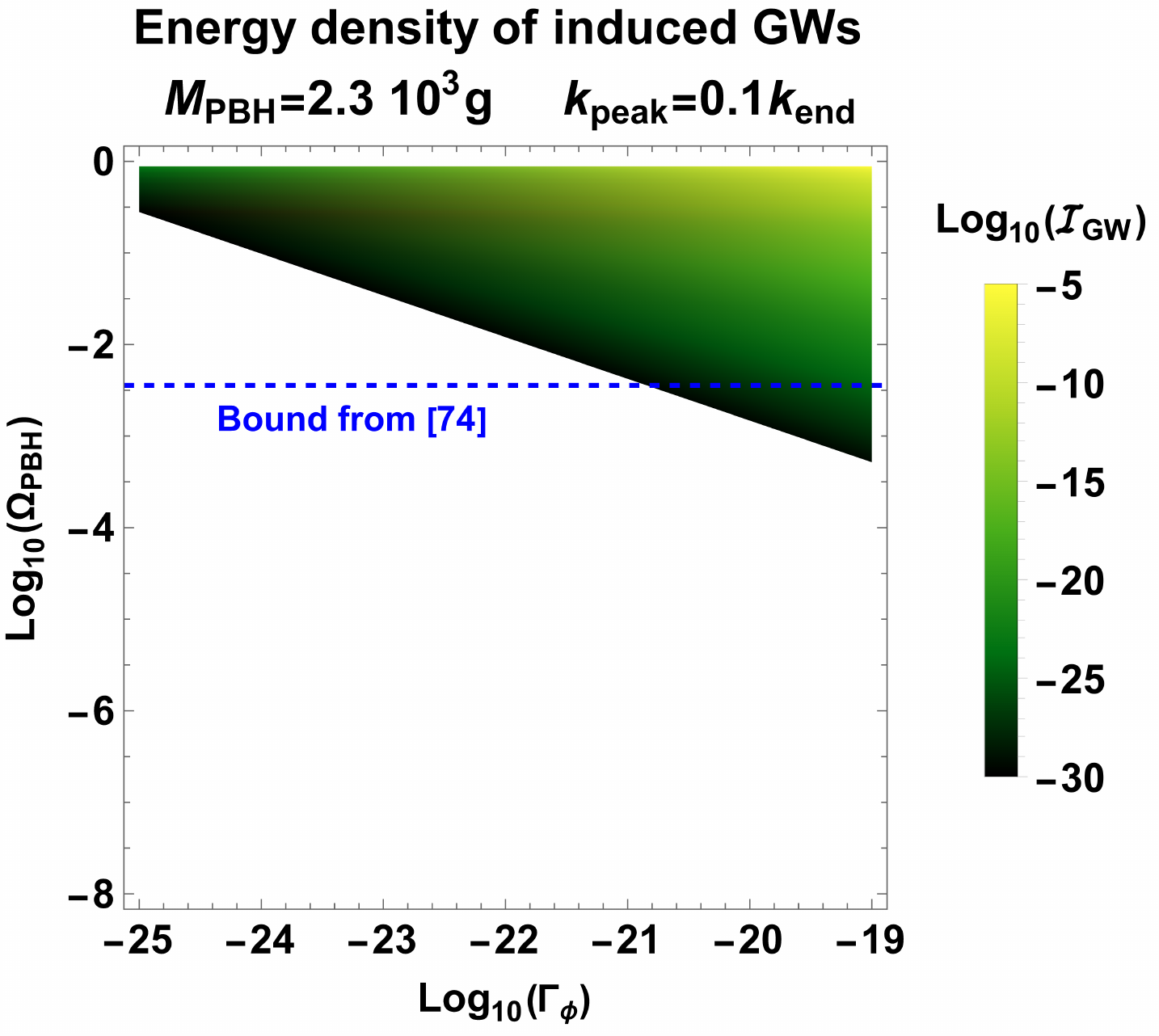}
         \caption{}
         \label{fig:GWs1}
     \end{subfigure}
        \caption{Total energy density of SIGWs ($I_{GW}$) for \textbf{(a)} $k_\text{peak}=10^{-2}k_\text{end}$ and \textbf{(b)} $k_\text{peak}=10^{-1}k_\text{end}$ as a function of the fractional energy density of PBH ($\Omega_{\rm PBH}$) at formation and the decay rate of the field ($\Gamma_\phi$). The dashed blue line corresponds to the bound from \cite{Papanikolaou:2020qtd}, given in eqn.~\eqref{eq:bound}, which does not consider the effects of the decay rate. It corresponds to the maximum initial fractional energy density of PBHs that does not overproduce GWs and reaches the BBN bound. By considering the effect of the decay rate, we extend this bound to higher values of $\Omega_{\text{PBH}}$.}
        \label{fig:GWs}
\end{figure}
Fig.~\ref{fig:GWs} shows the computation of $\mathcal I_\text{GW}$ for the two values of $k_\text{peak}$ considered in this work and as a function of the decay rate of the scalar field $\Gamma_\phi$ and the initial fractional energy density of PBH, $\Omega_\text{PBH}(t_\text{in})$. This last, as shown in App.~\ref{sec:appendix2}, depends directly on $\mathcal A_\text{peak}$ and $\sigma$ (see eqns.~\eqref{eq:Omega-small-sigma}, \eqref{eq:Omega-high-sigma}, and \eqref{eq:estimation-Omega}). To reproduce this plot we have selected the values of $\mathcal I_\text{GW}$ that satisfy the BBN bound \eqref{eq:BBN} and also the cases where PBH dominate before their evaporation (otherwise the production of GWs is not possible through this mechanism). Further, as explained in Sec.~\ref{sec:inflation-and-preheating}, we consider the cases where the temperature of radiation is higher than the evaporation temperature of the PBH. This effect mainly translates into a lower $k_\text{eva}$ which, looking at \eqref{eq:energy-density-final}, induces a higher amount of SIGWs, since PBH dominate for a longer time. However, this scenario does not affect too much the production of SIGWs. Also, for comparison, the blue dashed line corresponds to the bound shown in \cite{Papanikolaou:2020qtd}, which translated to our notation is given by
\begin{equation}\label{eq:bound}
    \Omega_\text{PBH}(t_\text{in})<1.4\times 10^{-4}\left(\frac{10^9\text{g}}{ M_{\text{PBH},0}}\right)^{1/4}.
\end{equation}
As it can be seen, by considering an initial matter-dominated universe together with a decay rate of the scalar field into radiation, one can relax the constraints on the initial abundance of PBH at production time. This means that high initial values of $\Omega_\text{PBH}$ are allowed if the PBH dominate for a short period, which is possible if $\Gamma_\phi$ is small. On the contrary, a small initial abundance of PBH needs more time to reach the BBN bound and overproduce GWs, that is, small $\Omega_\text{PBH}$ and high $\Gamma_\phi$. 

\begin{figure}
     \centering
     \begin{subfigure}[b]{0.49\textwidth}
         \centering         \includegraphics[width=\textwidth]{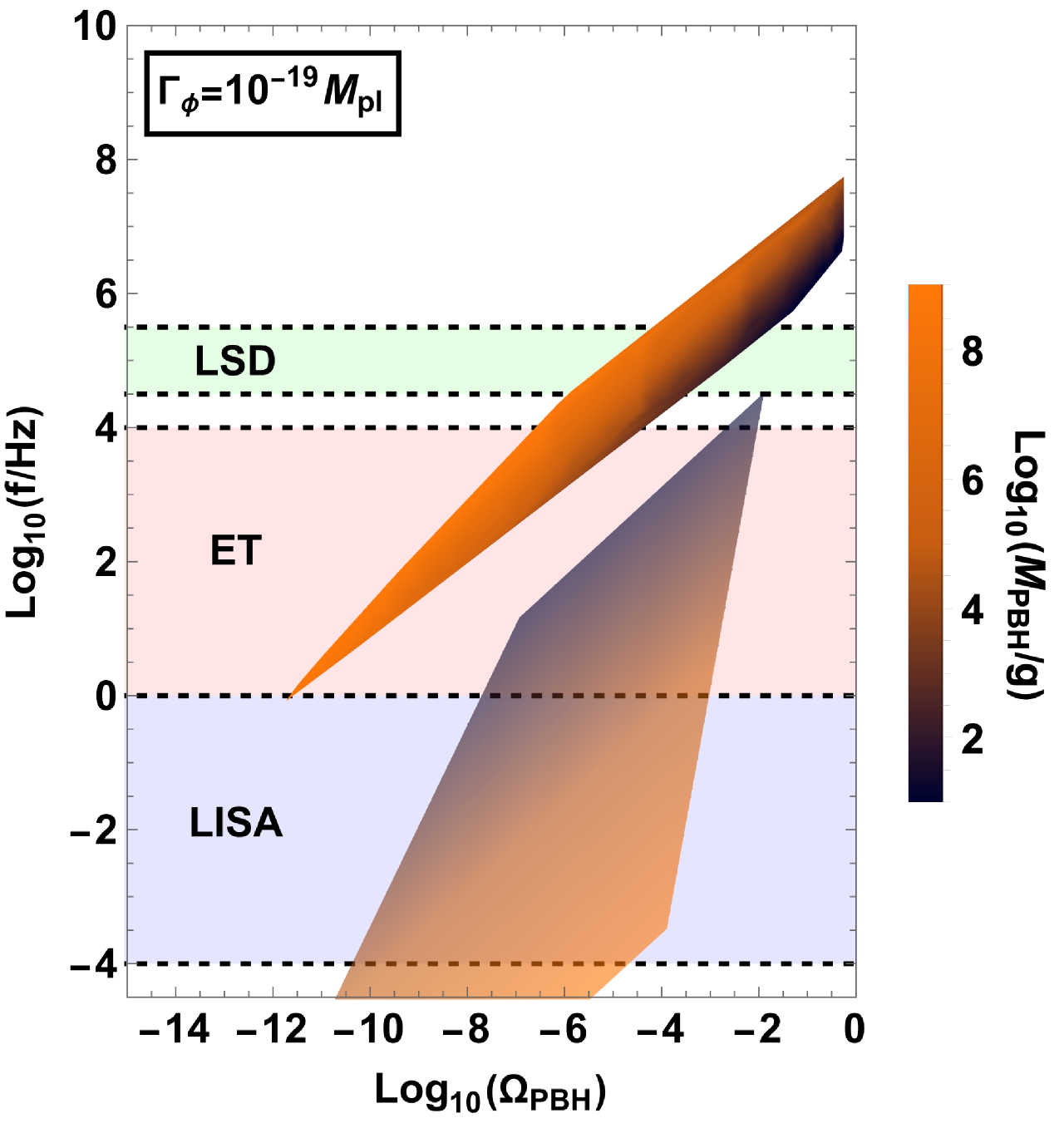}
         \caption{}
         \label{fig:large-decay-rate}
     \end{subfigure}
     \begin{subfigure}[b]{0.49\textwidth}
         \centering         \includegraphics[width=\textwidth]{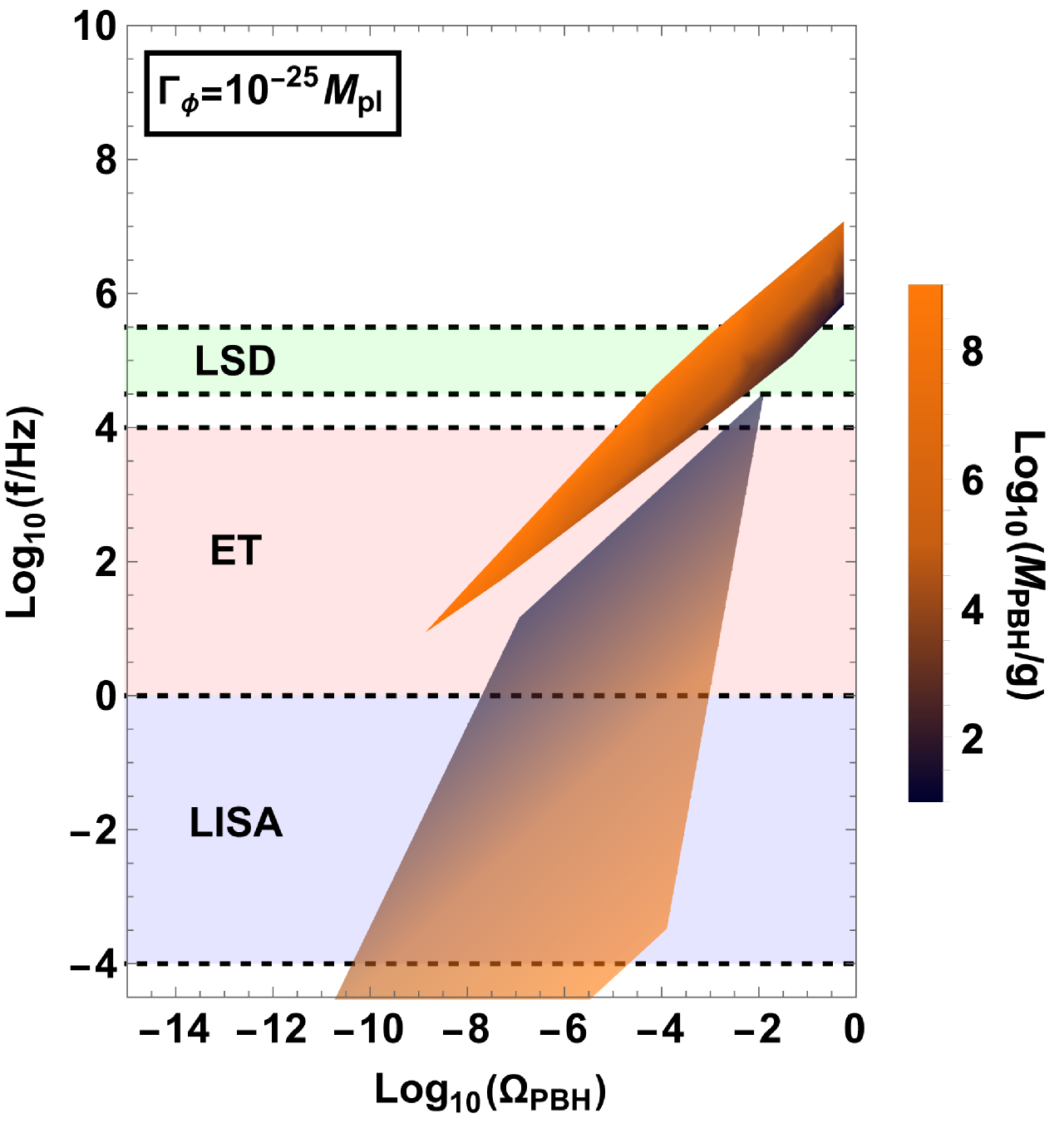}
         \caption{}
         \label{fig:small-decay-rate}
     \end{subfigure}
        \caption{Mass of the PBH as a function of the frequency at which the fractional energy density of GWs peaks and the initial fractional energy density of PBH. The green, pink, and purple bands correspond to the range of detectable frequencies of the LSD, ET, and LISA GW detectors, respectively. {Brighter orange-to-black colors correspond to our results, whereas lighter colors correspond to the results of ref.~\cite{Papanikolaou:2020qtd}.}}
        \label{fig:decay-rate}
\end{figure}

In Fig.~\ref{fig:decay-rate}, we show the mean mass of the PBH distribution as a function of the fractional density of PBH and the frequency of the peak of the SIGWs produced in each case. To compare, we also show the frequency ranges of some planned GW detectors, such as the Levitated Sensor Detector (LSD) \cite{Aggarwal:2020umq}, the Einstein Telescope (ET) \cite{Maggiore:2019uih}, and the Large Interferometer Space Antenna (LISA) \cite{Amaro-Seoane:2017ekt}. This reveals that, for some regions of the parameter space, the frequency of the GWs falls into the detectable range of the LSD and ET detectors. However, their sensitivity is insufficient to detect these GWs, which further motivates their refinement. In addition, we also show the effect of changing the decay rate. A larger decay rate (left plot) implies that PBH have more time to dominate, as the scalar field decays sooner. As a consequence, GWs are produced more abundantly and on a wider span of frequencies. On the contrary, a smaller decay rate (right plot) implies a reduced production of GWs. To produce this plot, we have considered the PBH that have enough time to dominate ($t_\text{PBH}<t_\text{eva}$), and excluded the scenarios where GWs are overproduced ($\mathcal I_\Omega^\text{PBH}<\mathcal I_\text{BBN}$). This plot is aimed to be compared with Plot.~3 of \cite{Papanikolaou:2020qtd}, where the authors study the production of GWs from a PBH-dominated phase without focusing on the production mechanism or the inflationary details. {The results from Fig.~3 of \cite{Papanikolaou:2020qtd} are shown in lighter colors in Fig.~\ref{fig:decay-rate}. These, in general, correspond to lower frequencies and are placed below our results, shown in brighter colors.} In this work, in addition, we consider the whole evolution of the PBH, from formation mechanism to their domination and final evaporation, along with details of inflaton decay. Since the energy density of SIGWs $I_{\rm GW}$ relies on the PBH domination duration  (determined by the inflaton decay rate (through \eqref{eq:PBH-time}) and the PBH mass fraction (See \eqref{eq:KP-out-of-appendix} and \eqref{eq:Omega-PBH}), our results are significantly different from \cite{Papanikolaou:2020qtd}.  {Therefore, our considerations of PBH formation details and inflaton decay produce high-frequency GWs induced by PBH domination in contrast to the results in \cite{Papanikolaou:2020qtd}, which indicate a low-frequency domain.}
Note that the authors in \cite{Papanikolaou:2020qtd} evaluate the plot at the late matter-radiation equality, whereas Fig.~\ref{fig:decay-rate} is evaluated at the present time.

\begin{figure}
    \centering
    \includegraphics[width=0.78\linewidth]{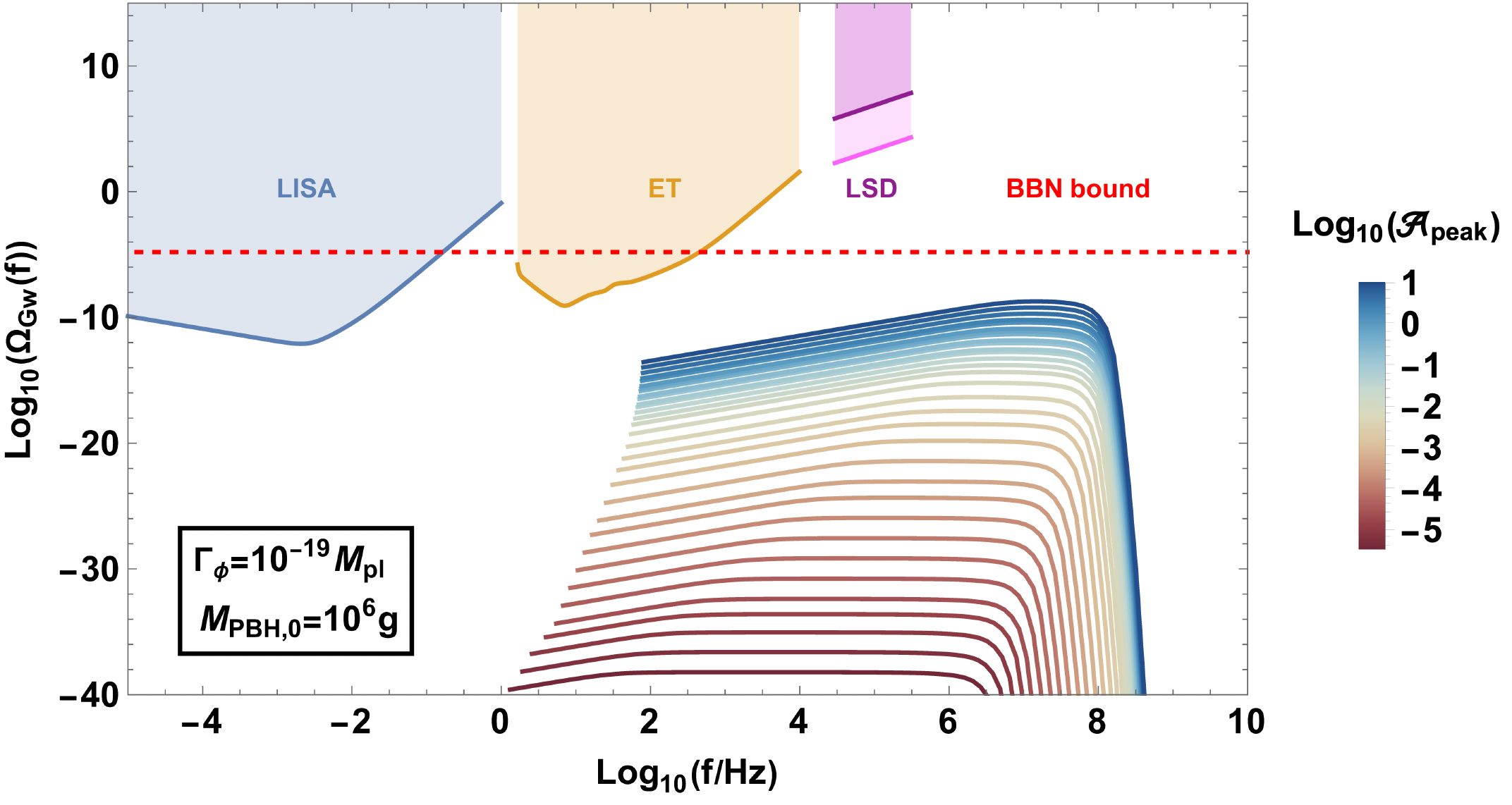}
    \caption{An example of GW spectra for the scenario where $\Gamma_\phi=10^{-19}\Mpl$ and $M_{\text{PBH},0}=10^6$g shown for several values of the amplitude of the Gaussian peak $\mathcal{A}_\text{peak}$.}
    \label{fig:GW-spectrum}
\end{figure}

Finally, and for completeness, we show in Fig.~\ref{fig:GW-spectrum} the GW spectrum as a function of the frequency for the case of a decay rate of $\Gamma_\phi=10^{-19}\Mpl$ and a mean PBH mass of $M_{\text{PBH},0}=10^{6}$g. The result is shown for several values of $\mathcal{A}_\text{peak}$, which translates into different mass fractions of PBH, $\Omega_{\text{PBH}}(t_{\text{in}})$. As explained above, the GW spectrum lies below the sensitivity of the future GW detectors. However, it is worth mentioning that one could also consider detectors based on electromagnetic resonant cavities~\cite{Herman:2022fau}. Several such detectors, both operational and planned, have been proposed \cite{Gatti:2024mde,Schneemann:2024qli}, offering very high experimental sensitivities. Nevertheless, their characteristic frequency ranges lie in the GHz region of the spectrum, slightly above the upper limit of the predicted signal in our scenario.

%%%%%%%%%%%%%%%%%%%%%%%%%%%%%%%%%
%% CONCLUSIONES
%%%%%%%%%%%%%%%%%%%%%%%%%%%%%%%%%

\section{Conclusions}\label{sec:conclusions}

In this work, we have explored the formation and evolution of PBH in an early matter-dominated universe, focusing on their potential dominance, which provides an alternative reheating mechanism through Hawking radiation. Our approach focuses on the Khlopov-Polnarev formalism \cite{Khlopov:1980mg,Khlopov:1982ef,Polnarev:1985btg,Khlopov:2008qy,Polnarev:1981,Carr:2020gox,Carr:2021bzv,Harada:2016mhb} to describe PBH formation during the preheating (matter-dominated) phase, considering an extended distribution of perturbations rather than a monochromatic one. This scenario differs significantly from the standard ones considered in the literature \cite{Papanikolaou:2020qtd,Domenech:2023jve,Domenech:2024wao,Domenech:2020ssp,Papanikolaou:2022chm,Papanikolaou:2024kjb,He:2024luf}, where PBH are considered to form during radiation domination and the mass fraction is computed using the Press-Schechter formalism \cite{Press:1973iz}. To achieve a significant amount of PBH, we consider a Starobinsky-like inflationary model \cite{Starobinsky:1980te} with some feature in the potential that amplifies perturbations around a particular scale, parametrized with a Gaussian peak, eqn.~\eqref{eq:power-spectrum-total}. PBH domination, in our framework, is affected by the inflaton field decay into radiation via a decay rate $\Gamma_\phi$, whose values we choose according to the Starobinsky model \cite{Jeong:2023zrv}. Then, the evolution of the energy densities is solved with a system of coupled Boltzmann equations \eqref{eq:Boltzmann-equations}. If PBH dominate for a sufficient amount of time (that depends on inflaton decay rate through \eqref{eq:PBH-time}), the Poissonian density fluctuations they produce induce GWs at second order in perturbation theory \cite{Baumann:2007,Ananda:2006,Assadullahi:2009} that could reach the BBN bound \eqref{eq:BBN}. Our study revealed that the duration of the PBH-dominated phase, and thus the production of GWs phase is highly sensitive to:
\begin{itemize}
    \item The decay rate of the inflaton field, $\Gamma_\phi$: A lower decay rate allows PBH to dominate for longer and induce more GWs.
    \item The fractional energy density of PBH, $\Omega_\text{PBH}$: If the PBH are produced more abundantly, they dominate sooner and induce more GWs. This quantity is directly related to $\mathcal A_\text{peak}$ and $\sigma$, see Appendix~\ref{sec:appendix2}.
    \item The mass of the PBH distribution, $M_{\text{PBH},0}$: The higher the mass, the later evaporation occurs and a longer PBH-dominated phase, which induces more GWs. This is inversely related to $k_\text{peak}$ \eqref{eq:mean-PBH-mass}.
\end{itemize}

We computed the fractional energy density of these induced GWs and found that the resulting signal could be within the detectable frequency range of future gravitational wave detectors such as the LSD and ET, see Fig.~\ref{fig:decay-rate}. However, as shown in Fig.~\ref{fig:GW-spectrum}, the experimental sensitivity of these detectors is insufficient to detect the computed signal, highlighting the need for further improvements or the design of new instruments in this aspect. Further, our results indicate that considering an early matter-dominated phase together with a decay rate for the inflation field allows for a relaxation of earlier constraints on PBH, as Fig.~\ref{fig:GWs} reveals. This suggests that PBH could have played a more significant role in cosmic evolution than early studies indicate,  which highlights the importance of considering the interplay between PBH formation during a matter-dominated phase using the Khlopov-Polnarev formalism, the inflaton decay to radiation, and the emission of induced gravitational waves in the early universe.

%%%%%%%%%%%%%%%%%%%%%%%%
%% AKNOWLEDGEMENTS
%%%%%%%%%%%%%%%%%%%%%%%%

\acknowledgments
Daniel del-Corral is grateful for the support of grant UI/BD/151491/2021 from the Portuguese Agency Funda\c{c}\~ao para a Ci\^encia e a Tecnologia. This research was funded by Funda\c{c}\~ao para a Ci\^encia e a Tecnologia grant number UIDB/MAT/00212/2020 and COST action 23130. KSK acknowledges support from the Royal Society Newton Fellowship Grant. The authors thank David Wands and Andrew Gow for useful discussions.

%%%%%%%%%%%%%%%%%%%%%%%%
%% APPENDIX A
%%%%%%%%%%%%%%%%%%%%%%%%

\appendix
\section{Mass fraction for high \texorpdfstring{$\sigma_k$}{TEXT}}\label{sec:appendix}

\begin{figure}
    \centering
    \includegraphics[width=0.75\linewidth]{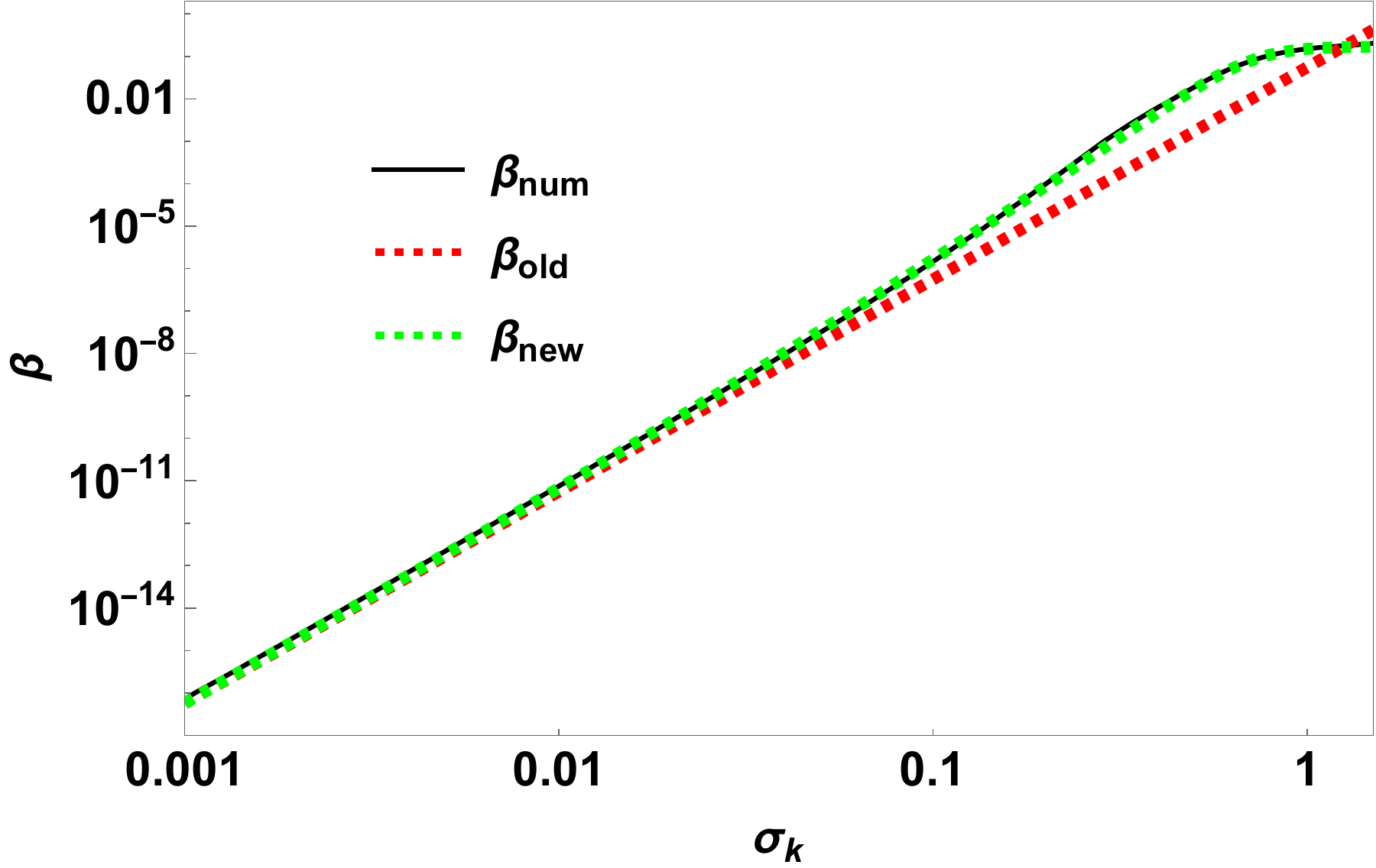}
    \caption{Mass fraction under the KP formalism as a function of the variance of the density perturbations $\sigma_k$. The curves labeled as $\beta_\text{num}$, $\beta_\text{old}$, and $\beta_{\text{new}}$ corresponds to eqns.~\eqref{eq:initial-mass-fraction}, $\beta(k)\simeq0.056\sigma_k^5$, and \eqref{eq:KP}, respectively.}
    \label{fig:mass-fraction}
\end{figure}

Following App.~B of \cite{Harada:2016mhb}, the mass fraction under the KP formalism is obtained by computing the following integral:
\begin{equation}\label{eq:initial-mass-fraction}
    \beta(k)=-\frac{675\sqrt{5}}{2\pi\sigma_k^6}\int_{-\frac12}^{\frac12}\dd u\,(2u-1)(2u+1)\int_{-1-\frac23u}^\infty\dd t\,\frac{2+2Az_*^2+A^2z_*^4}{A^3}e^{-Az_*},
\end{equation}
where only the anisotropy criterion has been considered and $A$ and $z_*$ are both functions of $(u,t)$, given by
\begin{subequations}\label{eq:A-and-z}
\begin{equation}
    A(t,u)=\frac92\left(\frac{t}{\sigma_k}\right)^2+10\left(\frac{u}{\sigma_k}\right)^2+\frac{15}2\left(\frac{1}{\sigma_k}\right)^2,
\end{equation}
\begin{equation}
    z_*(t,u)=\frac4\pi\left(t+\frac23u+1\right)^{-2}E\left(\sqrt{1-\left(u+\frac12\right)^2}\right),
\end{equation}
\end{subequations}
where $E(x)$ is the complete elliptic integral of the second kind, and $\sigma_k$ is the variance of the inflationary density perturbations, defined in \eqref{eq:variance}. Eqns.~\eqref{eq:initial-mass-fraction} and \eqref{eq:A-and-z} have been adapted to our notation. To recover the equations from App.~B of \cite{Harada:2016mhb} consider $\sigma_3^2=\frac{\sigma_k^2}{5}$.
The numerical solution of \eqref{eq:initial-mass-fraction} is shown in Fig.~\ref{fig:mass-fraction} as the continuous black curve labeled $\beta_\text{num}$ and as a function of the variance of the density perturbations $\sigma_k$.
Also shown in Fig.~\ref{fig:mass-fraction} is the analytical approximation $\beta(k)\simeq0.056\sigma_k^5$ for small $\sigma_k$ as the red dashed curve labeled $\beta_{\text{old}}$. One can observe that for $\sigma_k\gtrsim10^{-2}$, this analytical estimation deviates from the numerical solution, with one order of magnitude of deviation for $\sigma_k\sim\mathcal O(1)$. Since in this work, we study amplified perturbations, we find it useful to find a parametrization of the numerical solution valid also for high values of $\sigma_k$. This new parametrization is shown in Fig.~\ref{fig:mass-fraction} as the green dashed curve labeled as $\beta_{\text{new}}$, which fits better than $\beta_\text{old}$. It is given by
\begin{equation}\label{eq:KP}
    \beta_\text{new}(k)=\frac{A_1\,\sigma_k^5+A_2\,\sigma_k^6}{1+A_1\,\sigma_k^5+A_3\,\sigma_k^6},
\end{equation}
which asymptotes to the old estimation $\beta_\text{old}$ for small $\sigma_k$, and to the constant value $A_2/A_3$ for high $\sigma_k$. This last can be understood from the fact that strong anisotropy suppresses collapse and then higher $\sigma_k$ does not mean a higher $\beta(k)$. For instance, if a perturbation is highly anisotropic, \textit{i.e.}~very elongated, different regions of the perturbation will experience different gravitational forces and collapse at different rates. This leads to tidal shearing, since the perturbation stretches and deforms rather than forming a compact object, and to the formation of filaments or pancakes rather than PBH \cite{Khlopov:1980mg,Khlopov:1982ef,Polnarev:1985btg,Khlopov:2008qy,Polnarev:1981}. We numerically find the following values of the constants: $A_1=0.056$, $A_2=1.084$, and $A_3=6.536$. Eqn.~\eqref{eq:KP} offers, for the first time, a simple and rapid estimation of the mass fraction for the KP formalism over the whole range of $\sigma_k$ that avoids numerically-cost computations. This is the mass fraction used in this work to compute the abundance of PBH.

%%%%%%%%%%%%%%%%%%%%%%%%
%% APPENDIX B
%%%%%%%%%%%%%%%%%%%%%%%%

\section{Analytical solutions for \texorpdfstring{$\Omega_\text{PBH}(k)$}{TEXT}}\label{sec:appendix2}

\begin{figure}
    \centering
    \includegraphics[width=0.75\linewidth]{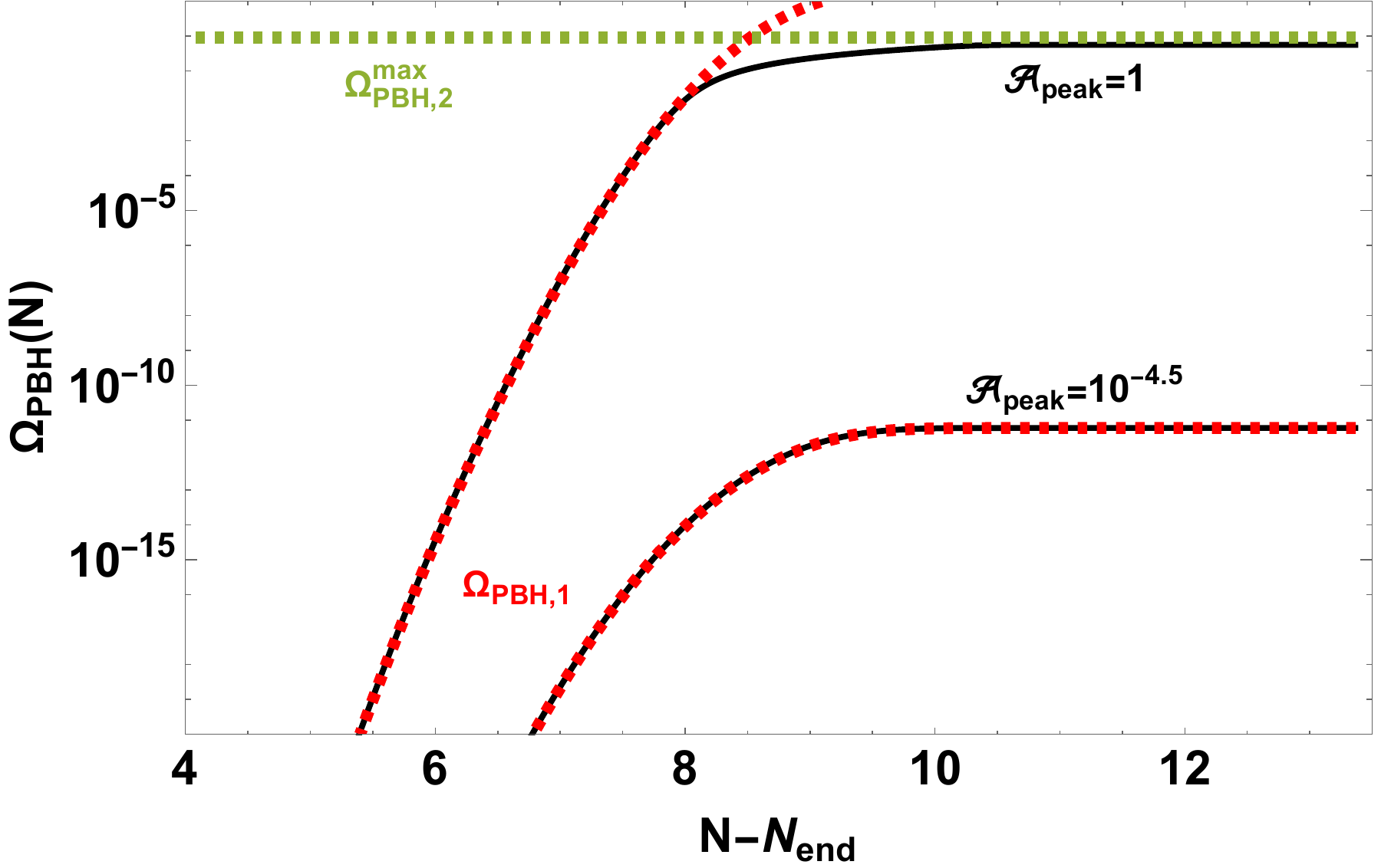}
    \caption{Fractional energy density of PBH as a function of the number of efolds from the end of inflation. The black curves show the numerical computation of \eqref{eq:Omega-PBH-k-appendix}, whereas the green and red curves the analytical approximation of \eqref{eq:estimation-small-amplification} and \eqref{eq:Omega-high-sigma}, respectively. The peak of the power spectrum is centered at $k_\text{peak}=10^{-2}k_\text{end}$.}
    \label{fig:analytic-app}
\end{figure}

We want to solve the integral
\begin{equation}\label{eq:Omega-PBH-k-appendix}
    \Omega_\text{PBH}(k)=3\int_k^{k_\text{end}}\beta(\tilde k)\dd\ln(\tilde k),
\end{equation}
where $\beta(k)$ is given by \eqref{eq:KP} in terms of $\sigma_k$. To simplify, let us consider that the power spectrum is given just by the Gaussian peak, that is, $\mathcal P_\mathcal{R}(k)\simeq\mathcal P_{\mathcal R}^\text{peak}(k)$. Now, using this in \eqref{eq:variance}, we obtain the following expression for the variance of the density perturbations
\begin{equation}\label{eq:variance-full}
    \sigma_k\simeq\frac{8\sqrt{\mathcal A_\text{peak}}}{5}\exp{\left[-\frac{\log{(k/k_\text{peak})^2}}{2\sigma^2}\right]}.
\end{equation}
To gain some insight into the behavior of $\Omega_\text{PBH}(k)$, we will solve analytically the integral \eqref{eq:Omega-PBH-k-appendix} in two regimes, small ($\sigma_k\lesssim1$) and large ($\sigma_k\gtrsim1$) variance. First, let us consider that the variance of the density perturbations is small for the whole range of $k$. In this case, we can safely consider 
\begin{equation}
\beta(k)\simeq A_1\sigma_k^5+A_2\sigma_k^6\qquad  (\sigma_k\lesssim1),
\end{equation}
and the integral \eqref{eq:Omega-PBH-k-appendix} reduces to the integral of sum of two Gaussians, that is
\begin{equation}\label{eq:Omega-PBH-x-1}
    \Omega_\text{PBH}(x)\simeq3\int_x^{x_\text{end}}\left(B_1e^{-\alpha_1x^2}+B_2e^{-\alpha_2x^2}\right)\dd x,
\end{equation}
where we have applied the change of variable $x=\ln(k/k_\text{peak})$ and defined
\begin{subequations}
    \begin{equation}
        B_1=A_1\left(\frac{8\sqrt{\mathcal A_\text{peak}}}{5}\right)^5,\quad  B_2=A_2\left(\frac{8\sqrt{\mathcal A_\text{peak}}}{5}\right)^6
    \end{equation}
    \begin{equation}
        \alpha_1=\frac{5}{2\sigma^2\ln(10)^2},\quad     \alpha_2=\frac{3}{\sigma^2\ln(10)^2}.
    \end{equation}
\end{subequations}
The integral of \eqref{eq:Omega-PBH-x-1} is now straightforward and gives
\begin{equation}\label{eq:estimation-small-amplification}
\begin{split}
    \Omega_{\text{PBH},1}(k,k_\text{end})&\simeq\frac{3\sqrt{\pi}}{2}\left\{\frac{B_1}{\sqrt{\alpha_1}}\left[\erf\left(\ln\left(\frac{k_\text{end}}{k_\text{peak}}\right)\sqrt{\alpha_1}\right)-\erf\left(\ln\left(\frac{k}{k_\text{peak}}\right)\sqrt{\alpha_1}\right)\right]\right.\\
    &+\left.\frac{B_2}{\sqrt{\alpha_2}}\left[\erf\left(\ln\left(\frac{k_\text{end}}{k_\text{peak}}\right)\sqrt{\alpha_2}\right)-\erf\left(\ln\left(\frac{k}{k_\text{peak}}\right)\sqrt{\alpha_2}\right)\right]\right\},
    \end{split}
\end{equation}
where $\erf(x)$ is the error function. Here we observe that, as the value of $k$ decreases (moving forward in time), the mass fraction reaches a constant value given by
\begin{equation}\label{eq:Omega-small-sigma}
    \Omega_{\text{PBH},1}^{\text{max}}=\frac{3\sqrt{\pi}}{2}\left(\frac{B1}{\sqrt{\alpha_1}}+\frac{B2}{\sqrt{\alpha_2}}\right).
\end{equation}
This occurs after the Gaussian peak has fully entered the horizon, when the production of PBH decreases drastically. The approximation \eqref{eq:estimation-small-amplification} is shown in Fig.~\ref{fig:analytic-app} in red as a function of the number of efolds from the end of inflation. It shows a good agreement with the numerical solution for small $\mathcal A_\text{peak}$.
However, as $\mathcal A_{\text{peak}}$ increases, a small portion of the Gaussian peak exceeds $\sigma_k\simeq1$ and for those modes $\beta(k)\neq A_1\sigma_k^5+A_2\sigma_k^6$. Instead, the mass fraction reaches a constant value 
\begin{equation}
    \beta(k)\simeq A_2/A_3\qquad(\sigma_k\simeq1).
\end{equation} 
Still, this approximation is only valid for the portion of the peak above the threshold value $\sigma_{\text{th}}=1$. Imposing $\sigma_k>\sigma_\text{th}$ in \eqref{eq:variance-full}, the following range of $k$ is obtained
\begin{equation}
    e^{-\sqrt{X(\sigma_\text{th})}}\lesssim \frac k{k_\text{peak}}\lesssim e^{\sqrt{X(\sigma_\text{th})}},
\end{equation}
where
\begin{equation}\label{eq:X-th}
    X(\sigma_\text{th})=-2\sigma^2\ln(10)^2\ln\left(\frac{5\sigma_\text{th}}{8\sqrt{\mathcal{A}_\text{peak}}}\right),
\end{equation}
which is only valid when $\sigma_\text{th}<\frac85\sqrt{\mathcal A_\text{peak}}$. In this interval, the integral \eqref{eq:Omega-PBH-k-appendix} is easily computed, and gives
\begin{equation}\label{eq:Omega-high-sigma}
    \Omega_{\text{PBH},2}^\text{max}\simeq\frac{6A_2}{A_3}\sqrt{X(\sigma_{\text{th}})}.
\end{equation}
This is by definition the highest contribution to $\Omega_\text{PBH}$, and thus serves as an upper bound on the mass fraction of PBH. It is shown in Fig.~\ref{fig:analytic-app} in green for the case when \eqref{eq:X-th} is valid, that is, when the Gaussian peak is above $\sigma_k>1$, and shows also good agreement with the numerical solution. In short, the maximum abundance of PBH for our model can be estimated as follows
\begin{equation}\label{eq:estimation-Omega}
    \Omega_\text{PBH}^\text{max}=\text{min}\left\{\Omega_{\text{PBH},1}^{\text{max}},\Omega_{\text{PBH},2}^{\text{max}}\right\}.
\end{equation}

%%%%%%%%%%%%%%%%%%%%%%%%
%% APPENDIX C
%%%%%%%%%%%%%%%%%%%%%%%%

\section{Analytical solutions for \texorpdfstring{$\Omega_\text{GW}^\text{PBH}$}{TEXT}}\label{sec:appendix3}

\begin{figure}
    \centering
    \includegraphics[width=0.75\linewidth]{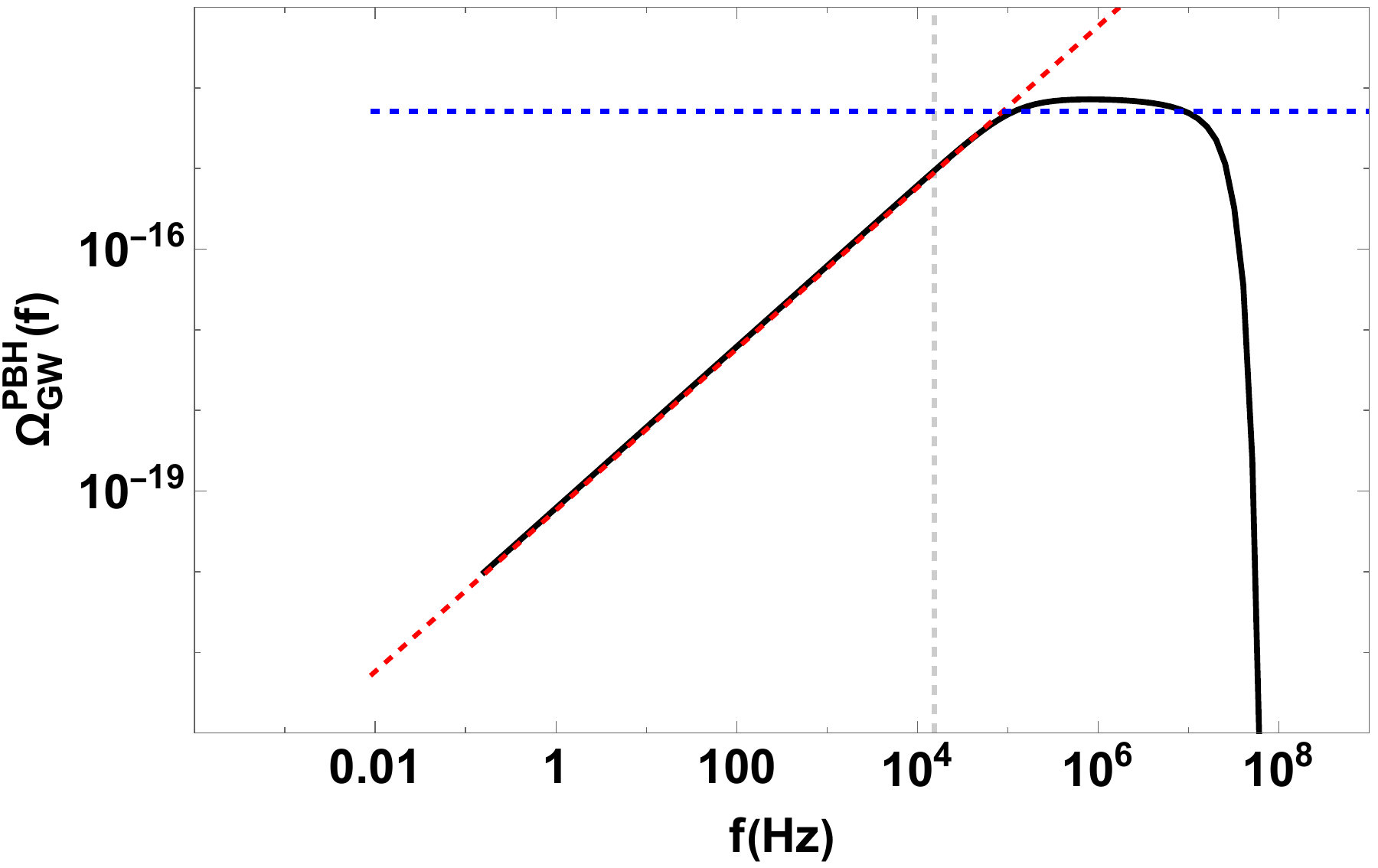}
    \caption{Numerical solution (continuous black) of the fractional energy density of SIGWs, eqn.~\eqref{eq:Omega-full} and the analytical approximations for small (large) $k/k_\text{PBH}$ in red (blue). The vertical gray dashed line corresponds to the frequency associated with the mode $k_\text{PBH}$. The set of parameters chosen is: $k_\text{peak}=0.01k_\text{end}$, $\mathcal A_\text{peak}=0.1$, and $\Gamma_\phi=4.5\times10^{-18}$.}
    \label{fig:Omega-app}
\end{figure}

The fractional energy density of SIGWs is given by eqn.~\eqref{eq:energy-density-final} as follows:
\begin{equation}\label{eq:Omega-full}
\begin{split}
    \Omega_{\text{GW}}^{\text{PBH}}(k)&\simeq\frac{16\,\Omega_\gamma^0}{27\pi^2}\frac{k}{k_{\text{eva}}^2k_\text{UV}^6}\int_{k_\text{eva}}^{k_\text{UV}}\dd\tilde k\int_{-1}^1\dd\mu\,\tilde k^6(1-\mu^2)^2\\
    &\cdot\left[5+\frac49\left(\frac{\tilde k}{k_\text{PBH}}\right)^2\right]^{-2}\left[5+\frac49\left(\frac{|\bm k-\tilde{\bm k}|}{k_\text{PBH}}\right)^2\right]^{-2},
    \end{split}
\end{equation}
where we have substituted the power spectrum of SIGWs \eqref{eq:power-SIGWs} and the power spectrum of PBH density fluctutations \eqref{eq:power-spectrum-density-PBH-final}. Further, we consider $g(k,t_\text{eva})\simeq1$ for the modes of interest. We give now analytical estimations of \eqref{eq:Omega-full} to understand the dependence of $\Omega_\text{GW}^\text{PBH}$ with the parameters of the model. Defining
\begin{equation}\label{eq:definition-x-y}
    x\equiv\frac{\tilde k}{k_\text{PBH}},\qquad  y\equiv\frac{k}{k_\text{PBH}},
\end{equation}
the integral in \eqref{eq:Omega-full} is simplified to
\begin{equation}
\begin{split}
    \Omega_{\text{GW}}^{\text{PBH}}(y)&\simeq\frac{16\,\Omega_\gamma^0}{27\pi^2}\frac{k_\text{PBH}^8}{k_{\text{eva}}^2k_\text{UV}^6}y\\
    &\int_{x_\text{eva}}^{x_\text{UV}}\dd x\int_{-1}^1\dd\mu\, x^6\left(5+\frac49x^2\right)^{-2}\left(5+\frac49(x^2+y^2-2xy\mu)\right)^{-2},
    \end{split}
\end{equation}
where $x_\text{eva}$ and $x_\text{UV}$ are defined using \eqref{eq:definition-x-y}. The main contribution to the integral in $\mu$ comes from $\mu=0$, which further simplifies the integral to
\begin{equation}
    \Omega_\text{GW}^\text{PBH}(y)\simeq\frac{16\,\Omega_\gamma^0}{27\pi^2}\frac{k_\text{PBH}^8}{k_{\text{eva}}^2k_\text{UV}^6}y\int_{x_\text{eva}}^{x_\text{UV}}\dd x \,\upsilon(x,y),
\end{equation}
where $\upsilon(x,y)$ is defined as
\begin{equation}
    \upsilon(x,y)= x^6\left(5+\frac49x^2\right)^{-2}\left(5+\frac49(x^2+y^2)\right)^{-2}.
\end{equation}
A primitive of $\upsilon(x,y)$ with respect to the variable $x$ is given by
\begin{equation}
\begin{split}
    \Upsilon(x,y)=\frac{6561}{4096y^6}
&\left[-\frac{4050 x y^2}{45 + 4 x^2} - \frac{2 x y^2 (45 + 4 y^2)^2}{45 + 4 x^2 + 4 y^2} + 675 \sqrt{5} (9 + y^2) \tan^{-1} \left(\frac{2x}{3\sqrt{5}}\right)\right.\\&\left. + (y^2-45) (4 y^2+45)^{\frac{3}{2}} \tan^{-1} \left(\frac{2x}{\sqrt{45 + 4 y^2}}\right) \right].
\end{split}
\end{equation}
Considering that $x_{\text{UV}}\gg x_\text{eva}$, we have that
\begin{equation}
    \int_{x_\text{eva}}^{x_\text{UV}}\dd x\,\upsilon(x,y)=\Upsilon(x_\text{UV},y)-\Upsilon(x_\text{eva},y)\simeq\Upsilon(x_\text{UV},y).
\end{equation}
Now, for the modes that enter the horizon during PBH domination ($y\ll1$), and for a very large $x_\text{UV}$, the function $\Upsilon(x_\text{UV},y)$ asymptotes to
\begin{equation}
    \Upsilon(x_\text{UV}\gg1,y\ll1)\simeq\Upsilon(\infty,0)=\frac{2187\sqrt{5}\pi}{4096}.
\end{equation}
On the contrary, for the modes that are already inside the horizon during PBH domination ($y\gg1$), and again for a very large UV cut-off, the function $y\Upsilon(x_\text{UV},y)$ asymptotes to the value
\begin{equation}
    y\Upsilon(x_\text{UV}\gg1,y\gg1)\simeq y\Upsilon(\infty,\infty)=\frac{6561\pi}{1024}.
\end{equation}
Then, the fractional energy density of SIGWs can be estimated as
\begin{equation}\label{eq:Omega-small-k}
    \Omega_\text{GW}^\text{PBH}(k\ll k_\text{PBH})\simeq\frac{81\sqrt{5}\,\Omega_\gamma^0}{256\pi}\frac{k\, k_\text{PBH}^7}{k_{\text{eva}}^2k_\text{UV}^6}\sim k
\end{equation}
for modes entering the horizon when PBH dominate and as
\begin{equation}\label{eq:Omega-large-k}
    \Omega_\text{GW}^\text{PBH}(k\gg k_\text{PBH})\simeq\frac{243\,\Omega_\gamma^0}{64\pi}\frac{k_\text{PBH}^8}{k_{\text{eva}}^2k_\text{UV}^6}\sim \text{cte},
\end{equation}
for the modes that are already inside the horizon when the PBH starts dominating. This last serves us as an estimation of the maximum amount of GWs produced. Eqns.~\eqref{eq:Omega-small-k} and \eqref{eq:Omega-large-k} are shown in Fig.~\ref{fig:Omega-app} in dashed red and blue, respectively, and for a particular choice of parameters, see the caption for details. We observe that both agree well with the full numerical solution of \eqref{eq:Omega-full}, shown in continuous black line.

%%%%%%%%%%%%%%%%%%%%%%%%
% BIBLIOGRAPHY
%%%%%%%%%%%%%%%%%%%%%%%%

\bibliographystyle{JHEP.bst}
\bibliography{BIBLIO.bib}

\end{document}